\documentclass{emulateapj}

\newcommand{\yt}{\texttt{yt}}

\usepackage{natbib}
\usepackage{graphicx}
\usepackage{hyperref}

\begin{document}


\title{yt: A Multi-Code Analysis Toolkit for Astrophysical Simulation Data}
\author{Matthew J.~Turk\altaffilmark{1}, 
Britton D. Smith\altaffilmark{2},
Jeffrey S.~Oishi\altaffilmark{3,4},
Stephen Skory\altaffilmark{5},
Samuel W.~Skillman\altaffilmark{5,6},
Tom Abel\altaffilmark{3,4},
Michael L.~Norman\altaffilmark{1}
}
\altaffiltext{1}{Center for Astrophysics and Space Science, University of
California, San Diego}
\altaffiltext{2}{Department of Physics and Astronomy, 
Michigan State University, East Lansing, MI}
\altaffiltext{3}{Kavli Institute for Particle Astrophysics and Cosmology,
Stanford University, 2575 Sand Hill Road, Menlo Park, CA 94025}
\altaffiltext{4}{Department of Physics,
Stanford University, Stanford, CA 94305}
\altaffiltext{5}{Center for Astrophysics and Space Astronomy,
University of Colorado, Boulder}
\altaffiltext{6}{DOE Computational Science Graduate Fellow}

\begin{abstract}
The analysis of complex multiphysics astrophysical simulations presents a unique
and rapidly growing set of challenges: reproducibility, parallelization, and
vast increases in data size and complexity chief among them. In order to meet
these challenges, and in order to open up new avenues for collaboration between
users of multiple simulation platforms, we present \yt{}\footnote{Available at
\url{http://yt.enzotools.org/}}, an open source, community-developed
astrophysical analysis and visualization toolkit.  Analysis and visualization
with \yt{} are oriented around physically relevant quantities rather than
quantities native to astrophysical simulation codes.  While originally designed
for handling Enzo's structure adaptive mesh refinement (AMR) data, \yt{} has
been extended to work with several different simulation methods and simulation
codes including Orion, RAMSES, and FLASH.  We report on its methods for
reading, handling, and visualizing data, including projections, multivariate
volume rendering, multi-dimensional histograms, halo finding, light cone
generation and topologically-connected isocontour identification.  Furthermore,
we discuss the underlying algorithms \yt{} uses for processing and visualizing
data, and its mechanisms for parallelization of analysis tasks.
\end{abstract}
\keywords{methods: data analysis; methods: numerical; methods: n-body
simulations; cosmology: theory;}
\maketitle

\section{Introduction}
 In the last decade, multiphysics astrophysical simulations have increased
 exponentially in both sophistication and size \citep{
2005Natur.435..629S, 
2008JPhCS.125a2008K, 
2009JCoPh.228.6833R, 
2008MNRAS.390.1326O, 
2007arXiv0705.1556N, 
2010arXiv1008.4368K, 
2010arXiv1008.2801A, 
2010arXiv1002.3660K, 
2000ApJS..131..273F, 
2007ApJ...659L..87A}; 
however, the software tools to mine those simulations have not kept pace.
Typically, methods for examining data suffer from a lack of agility,
discouraging exploratory investigation. Massively parallel visualization tools
such as VisIT and ParaView \citep{visit_paper, paraview_paper}, are quite
general, serving the needs of many disparate communities of researchers. While
a multi-code, astrophysical analysis system could be built on top of one of
these tools, we have chosen a lighter weight approach that we feel is
complementary.  The lack of domain-specific quantitative analysis tools
designed for astrophysical data leads to the development of specialized tools
by individual researchers or research groups, most of which are never shared
outside the research group. This can substantially inhibit collaboration
between different groups--even those using the same simulation code.

Furthermore, tools developed by a single research group are often tightly
coupled to a specific simulation code or project.  This results in a constant
process of reinvention: individual research groups create analysis scripts
specific to a single simulation tool that read data from disk, assemble it in
memory, convert units, select subsections of that data, perform some type of
quantitative analysis and then output a reduced data product. When
collaborative analysis between research groups exists, it often includes
creation of intermediate data formats, requiring substantial ``last mile''
efforts to ensure correct units, format, and other data-transport details.

This fractionation of the astrophysical community demonstrates a clear need for
a flexible and cross-code software package for quantitative data analysis and
visualization.  In this paper we present \yt{}, a data analysis and
visualization package that works with several astrophysical simulation codes.
\yt{} is developed openly and is freely available at
\url{http://yt.enzotools.org/}.  It has been designed to be a common
platform for simulation analysis, so that scripts can be shared across groups
and analysis can be repeated by independent scientists.  Historically, \yt{}
was initially developed to examine slices and projected regions through deeply
nested adaptive mesh refinement cosmological simulations conducted with Enzo
\citep{bryan97,oshea04}, but it was quickly repurposed to be a multi-code
mechanism for data analysis and visualization.

By making this tool available, we hope not only to encourage cross-group
collaboration and validation of results, but to remove or at least greatly
lower the barrier to entry for exploratory simulation analysis.  \yt{} provides
mechanisms for conducting complete analysis pipelines resulting in publication
quality figures and data tables, as well as the necessary components for
constructing new methods for examining data.  The concepts for data handling
and representation in \yt{} are certainly not new, but their application to
astrophysical data enables complex, detailed analysis pipelines to be shared
between individuals studying disparate phenomena using disparate methods.  This
enables and even encourages reproducibility and independent verification of
results.

\yt{} is primarily written in Python\footnote{\url{http://www.python.org/}}
with several core routines written in C for fast computation.  \yt{} heavily
utilizes the NumPy library
\citep[][\url{http://numpy.scipy.org}]{numpy_paper}, and is itself a Python
module suitable for direct scripting or access as a library.  A community of
users and developers has grown around the project, and it has been used in
numerous published papers and posters \citep[For example][]{
2010arXiv1008.2402C, 2010ApJ...715.1575S, 2010arXiv1006.3559S,
2010MNRAS.tmp.1530A, 2010ApJ...721.1105B, 2009ApJ...698.1795H,
2009ApJ...694L.123K, 2009Sci...325..601T, 2008Offnerposter,
2008MNRAS.385.1443S}.

In order to accomodate the diverse computing environments on which
astrophysical simulations are run, \yt{} was designed to use primarily
off-screen rendering and scripting interfaces, although several smaller tools
are provided for specific, interactive visualization tasks.  The former method
is well-suited to remote visualization and can be run via a job execution queue
on a batch-compute cluster, such as those on which the underlying simulation
are run.  \yt{} is subdivided into several sub-packages for data handling, data
analysis, and plotting.  This modularity encourages the creation of reusable
components for multi-step analysis operations, \emph{which can then be used
without modification on data from any simulation code \yt{} supports}.

\begin{figure}[ht]
\includegraphics[width=0.35\textwidth]{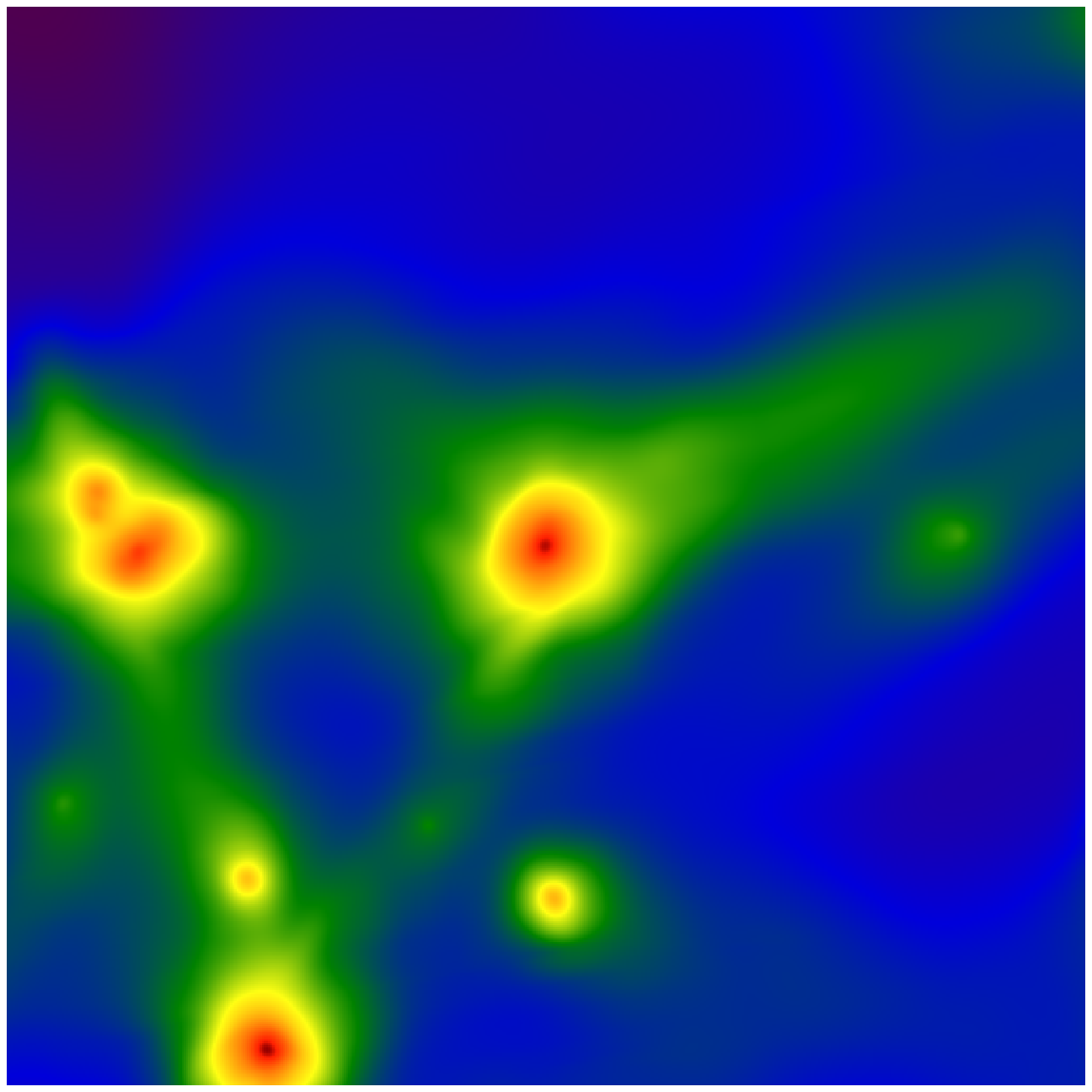}
\begin{verbatim}
from yt.mods import *
from yt.visualization.volume_rendering.api import \
    ProjectionTransferFunction
from yt.visualization.image_writer import \
    write_image

pf = load("DD1701/DD1701")

halos = HaloFinder(pf)

tf = ProjectionTransferFunction()

for halo in halos:
    center = halo.center_of_mass()
    radius = halo.maximum_radius()
    sp = halo.get_sphere()
    L_vec = sp.quantities["AngularMomentumVector"]()
    cam = pf.h.camera(center, L_vec, 10.0*radius,
                      resolution=(512, 512),
                      log_fields = False,
                      transfer_function=tf)
    im = cam.snapshot()
    write_image(na.log10(im), "halo_%04i.png" % halo.id)
\end{verbatim}
\caption{A script that loads data from disk, identifies halos within that dataset, and
then projects the density of those halos aligned with the angular momentum
vector of the halo.}\label{fig:halo_proj}
\end{figure}

As an example, in Figure~\ref{fig:halo_proj} we have included a script that
combines many of these components into a modifiable pipeline for visualization.
This script loads a dataset from disk (via \yt{}'s generic \texttt{load} command),
returning an instance of a Python class \texttt{StaticOutput}.  This object is
used as the source for a halo finding operation (\S~\ref{sec:halo_finding}),
which again returns an instance of a  Python class representing the collection
of identified halos.  Each halo's baryonic content is inspected individually
(\S~\ref{sec:data_containers}) and the angular momentum vector is calculated
(\S~\ref{sec:object_quantities}) and used as input to a volume rendering
operation (\S~\ref{sec:volume_rendering}).  Through this entire operation, the
underlying simulation data has largely been abstracted as a set of physical
objects.

In this paper, we will start by describing the mechanisms by which \yt{}
addresses and processes physical data (\S~\ref{sec:mechanisms}).  We then
discuss the visualization mechanisms available in \yt{}
(\S~\ref{sec:visualization}) such as projections, slices and volume rendering.
We proceed to enumerate the simulation codes \yt{} works with and discuss the
challenges they present (\S~\ref{sec:codes}).  In \S~\ref{sec:parallelism} we
describe the parallelism strategies used by \yt{}.  The mechanisms by which
\yt{} can be embedded in running simulations are presented and described in
\S~\ref{sec:embedding}.  Following this, we conclude by describing both the
pre-packaged analysis modules for \yt{} (\S~\ref{sec:analysis_modules}),
future directions in (\S~\ref{sec:future_directions}) and
discuss the community engagement and involvement around \yt{}
(\S~\ref{sec:conclusions}).

\section{Mechanisms for Interacting with Data}\label{sec:mechanisms}

The vast majority of adaptive mesh refinement calculations in the astrophysical
literature are computed on a rectilinear grid; while this affords a number of
computational efficiencies and conveniences, astrophysical phenomena as a whole
are not rectangular prisms and thus are ill-suited to analysis as rectangular
prisms.  This presents a fundamental disconnect between the data structures
utilized by simulations and the geoemetries found in nature.  Furthermore, the
task of selecting geometric regions in space requires substantial overhead.
\yt{} provides a number of convenience functions and mechanisms for addressing
data within astrophysical simulations that make the process of handling and
manipulating data straightforward.

The \yt{} codebase has been organized along several conceptual lines, each
corresponding to a set of tasks or classes in Python.  The primary mechanisms
for handling data are contained in the Python module \texttt{yt.data\_objects},
while all code and data structures specific to a particular simulation code
resides within a submodule of \texttt{yt.frontends} (such as
\texttt{yt.frontends.enzo}, \texttt{yt.frontends.orion}, etc).  To open a
dataset, the user creates an instance of a simulation code-specific subclass of
\texttt{StaticOutput}, a lightweight class that scans a parameter file and
obtains the necessary information to orient the dataset: the current time in
the simulation, the domain information, the mechanisms for converting units,
and the necessary file locations on disk.  A convenience function
(\texttt{load}) for automatically creating such an instance is provided, such
that it only requires a path on disk to the dataset of interest.  However,
geometric information about the manner in which data is laid out on disk or in
the simulation domain is compartmentalized to a \texttt{AMRHierarchy} object.
These objects are comparatively expensive to construct, as they contain a
hierarchy of \texttt{GridPatch} objects, all of which posses spatial and
parentage information.  These objects are not instantiated or constructed until
requested.  All data access is mediated by \texttt{AMRHierarchy} objects,
as noted below.

As an example, a sample \yt{} session where Enzo data is loaded off disk and
examined might look something like this:
\begin{verbatim}
>>> from yt.mods import *
>>> dataset = load("40-3D-3")
>>> print dataset.current_time
646.750
>>> print dataset.current_redshift
0.0
>>> dataset.hierarchy.print_stats()
level   # grids     # cells
---------------------------
  0          4        32768
  1         34       253496
  2        304       525784
----------------------------
           342       812048
\end{verbatim}
In this session, a relatively small dataset is loaded from disk.  The object
\texttt{dataset} contains a number of pieces of information about the
dataset: the current time, cosmological parameters (if applicable), the domain
size, and so on.  It is not until the attribute \texttt{.hierarchy} (or
\texttt{.h} for brevity) is accessed that \yt{} parses the underlying grid
patches that exist on disk, instantiates both the \texttt{AMRHierarchy} object,
and its component \texttt{GridPatch} objects, orients them in
space and sets up a mapping between grids and their location on disk.  All
further accessing of data, such as through data containers, is mediated by
the hierarchy object itself, rather than by the parameter file.

By relegating data handling to individual instances of classes, we
compartmentalize datasets; because each dataset is merely a variable, the
number that can be opened and simultaneously cross-compared is only limited by
the available memory and processing power of the host computer.  Furthermore,
datasets from different simulation codes can be opened and compared
simultaneously in memory.

\subsection{Data Containers}\label{sec:data_containers}

When handling astrophysical data, it is appropriate to speak of geometric
regions that outline the rough boundaries of physical objects: dark matter
halos as ellipsoids, protostars as spheres, spiral galaxies as cylinders, and
so on.  The central conceit behind \yt{} is the presentation to the user of a
series of physical objects with the underlying simulation largely abstracted.
For AMR data, this means hiding the language of grid patches, files on disk and
their interrelationships, and instead describing only geometric or physical
systems; these intermediate steps are handled exclusively by \yt{}, without
requiring any intervention on the part of the user.  For instance, to select a
spherical region, the user specifies a center and a radius and the underlying
\yt{} machinery will identify grid patches that intersect that spherical
region, identify which grid patches are the most highly-refined at all regions
within the sphere, locate the appropriate data on disk, read it and return this
data to the user.

The mechanisms in \yt{} for this abstraction are called \emph{data containers}.
These are Python objects, subclasses of a generic \texttt{AMRData} interface,
affiliated with a specific instance of an \texttt{AMRHierarchy} object, that
provide a consistent interface to a region of selected data.  This region can
be defined by geometric concerns or selected by criteria from physical
quantities in the domain.  Data contained in these objects is accessed in a
consistent manner and loaded on demand: the computational cost of creating a
box that covers an entire region is negligible, and until data is actually
accessed from that box the memory overhead remains likewise negligible.  By
abstracting the selection of and access to data in this manner, operations that
can be decomposed spatially or that are ``embarrassingly parallel" can be
transparently parallelized, without requiring the user's intervention.  (See
Section~\ref{sec:object_quantities} and Section~\ref{sec:parallelism}.) The
data containers implemented in \yt{} include spheres, rectangular prisms,
cylinders (disks), arbitrary regions based on logical operations,
topologically-connected sets of cells, axis-orthogonal and arbitrary-angle
rays, and both axis-orthogonal and arbitrary-angle slices.  Below, we show an
example of a hypothetical user accessing dataset \texttt{40-3D-3} (as shown
above), creating a sphere of radius $100~\mathrm{pc}$ centered at the origin,
and then accessing all ``Density'' values that reside within that sphere.
\begin{verbatim}
dataset = load("40-3D-3")
sphere = dataset.h.sphere( [0.0, 0.0, 0.0],
          100.0 / dataset["pc"] )
print sphere["Density"]
\end{verbatim}
When a data container is queried for a particular field (as in the final line
above), \yt{} will select the appropriate grid patches, read them from disk and
mask out regions where higher resolution data is available, and then return to
the user a one-dimensional array of values that constitute the data within a
region.  \yt{} also transparently allows for the creation of \emph{derived
fields}, fields that are defined as functions of the base fields output by the
simulation or even other derived fields.  These can be defined by the user and
supplied to \yt{}, and \yt{} provides a number of such fields.  For instance,
the user could define a derived field based on the density and temperature
present in the cell to estimate molecular hydrogen formation timescales, the
angular momentum with respect to a particular center, the total magnetic energy
in a cell, the spatial coordinates of a point, and so on.

Data containers provide several methods for data access.  The data can be
accessed directly, as in the above code listing, or through abstractions such
as \emph{object quantities}, described in Section~\ref{sec:object_quantities}.
Furthermore, data objects provide geometric information about the grid patches
that contribute to a given object, and through the usage of fields the total
mass, total volume and other physical quantities can be constructed.

Despite the pervasive abstraction of data selection, \yt{} also allows for
queries based on simulation data structures and access to raw fields in memory.
For instance, grid patch objects respect an identical protocol to data
containers. Accessing raw data in the terms the simulation code itself uses
allows \yt{} to be very useful during development and debugging (of both \yt{}
and the simulation code!)

The abstraction of data into data containers leads to the creation of systems
of components: data containers become ``sources'' for both analysis procedures
as well as visualization tasks.  These analysis procedures then become reusable
and the basis for chains of more complicated analysis tasks.  Using such
chains, a user can volume render a set of halos based on their angular momentum
vectors (as in Figure~\ref{fig:halo_proj}), color particles by merger history,
and even calculate disk inclination angles and mass fluxes.

\subsection{Data Fields}

Once a region of the simulation is selected for analysis, \yt{} must process
the raw data fields themselves. Its model for handling this data and processing
fundamental data fields into new fields describing \emph{derived quantities} is
built on top of an object model with which we can build automatically recursive
field generators that depend on other fields.  All fields, including derived
fields, are allowed to be defined by either a component of a data file, or a
function that transforms one or more other fields.  This indirection allows
multiple layers of definition to exist, encouraging the user to extend the
existing field set as needed.

By defining simple functions that operate via NumPy array operations,
generating derived fields is straightforward and fast.  For instance, a field
such as the magnitude of the velocity in a cell
\begin{equation}
   V = \sqrt{v_x^2 + v_y^2 + v_z^2}
\end{equation}
can be defined independently of the source of the data:
\begin{verbatim}
def VelocityMagnitude(field, data):
    return (data["x-velocity"]**2.0 +
            data["y-velocity"]**2.0 +
            data["z-velocity"]**2.0)**0.5
\end{verbatim}
Each operation acts on each element of the source data fields; this preserves the
abstraction of fields as undifferentiated sets of cells, when in fact those cells
could be distributed spatially over the entire dataset with varying cell
widths.

Once a function is defined, it is added to a global field container that
contains not only the fields, but a set of metadata about each field -- the
units of the field, the units of the field when projected, and any implicit or
explicit requirements for that field.  Field definitions can require that
certain parameters be provided (such as a height vector, a center point, a bulk
velocity and so on) or, most powerfully, that the data object has some given
characteristic.  This is typically applied to ensure that data is given in a
spatial context; for finite difference solutions, such as calculating the
gradient or divergence of a set of fields, \yt{} allows the derived field to
mandate that the input data provided in a three-dimensional structure.
Furthermore, when specifying that some data object be provided in three
dimensions, a number of buffer cells can be specified as well; the returned
data structure will then have those buffer cells taken from neighboring grids
(this utilizes \emph{covering grids}, as described in
\S~\ref{sec:fixed_resolution_grids}).  This enables higher-order methods to be
used in the generation of fields, for instance when a given finite difference
stencil extends beyond the computational domain of a single grid patch.  \yt{}
provides several fields that utilize buffer zones, such as the divergence of
the velocity and the spatially-averaged local density.

\subsection{Object Quantities}\label{sec:object_quantities}

In addition to the flexibility of defining field quantities at every point in
space, \yt{} provides the ability to examine quantities constructed from whole
regions in space.  These \emph{derived quantities} are available from any data
container present in \yt{}.  They are defined by some relationship of the
points contained within the container; this can be the bulk angular momentum
vector, the average velocity, the center of mass, the total mass, the moment of
inertia and so on.

These bulk quantities affiliated with objects are defined in two stages: the
calculation stage, wherein intermediate values can be created and stored, and a
reduction stage where the intermediate values are combined in some manner to
produce a final result.  This allows derived quantities to operate
transparently in parallel in an un-ordered fashion: a script that calculates
the total mass in a sphere occupying some volume in the simulation domain, when
run in parallel, will transparently distribute work (computation and disk IO)
between processors and then re-join the work to produce a final result.  This
parallelization process is described in more detail in
Section~\ref{sec:parallelism}.  For instance, the following script that uses
the \texttt{sphere} created in the above code listing, will return the
mass-weighted angular momentum vector of the baryonic content of that sphere:
\begin{verbatim}
L_vec = sphere.quantities[
    "AngularMomentumVector"]()
\end{verbatim}
The returned value (\texttt{L\_vec}) is a NumPy array and can be used in subsequent analysis.

These object quantities can be newly defined, can take any number of
parameters, and can take as input any derived field created by the user.  This
not only allows further flexibility on the part of the simulation, but allows
advanced, bulk manipulations of extremely large datasets to proceed in a
straightforward fashion.

\subsection{Fixed Resolution Grids}\label{sec:fixed_resolution_grids}

Multi-resolution data presents challenges to the application of certain classes
of algorithms, for example those using the Fast Fourier Transforms.  To address
this need, the creation of fixed-resolution (and three-dimensional) arrays of
data filling arbitrary rectangular prisms must be easy and accessible.  However,
unless the entire region under consideration is contained within a single grid
patch, it can be difficult to construct these arrays.  \yt{} creates these
arrays, or \emph{covering grids}, by an iterative process.  First all grids
intersecting or fully-contained within the requested rectangular prism are
selected.  These grids are then iterated over, starting on the coarsest level,
and used to fill in each point in the new array.  All grid cells that intersect
with the covering grid and where no finer-resolution data is available are
deposited into the appropriate cell in the \emph{covering grid}.  By this
method, the entire covering grid is filled in with the finest cells available to
it.  This can be utilized for generating ghost zones, as well as for new
constructed grids that span the spatial extent of many other grids that are
disjoint in the domain.

However, coarse cells are duplicated across all cells in the (possibly
finer-resolution) covering grid with which they intersect, which can lead to
unwanted resolution artifacts.  To combat this, a \emph{smoothed covering grid}
object is also available.  This object is filled in completely by iterating
over, from coarsest to finest, all levels $l < L$ where $L$ is the level at
which the covering grid is being extracted.  Once a given level has been filled
in, the grid is trilinearly interpolated to the next level, and then all new
data points from grids at that level replace existing data points.  We note,
however, that this does not explicitly perform the crack-fixing method described
in \cite{10.1109/VIS.2005.122}.  Nevertheless, it is suitable for
generating smoothed multi-resolution grids and constructing vertex-centered
data, as used in Section~\ref{sec:volume_rendering}.

\subsection{Multi-dimensional Profiles}

Distributions of data within the space of other variables are often necessary
when examining and analyzing data.  For instance, in a collapsing gas cloud or
galaxy cluster, examining the average temperature with increasing radius from a
central location provides a convenient means of examining the process of
collapse, as well as the effective equation of state.  To conduct this sort of
analysis, typically a multi-dimensional histogram is constructed, wherein the
values in every bin are weighted averages of some additional quantity.  In
\yt{}, the term \emph{profile} is used to describe any weighted average or
distribution of a variable with respect to an other independent variable or
variables.  Such uses include a probability density function of density, a
radial profile of molecular hydrogen fraction, and a radius, temperature, and
velocity phase diagram.  \yt{} provides functionality for these profiles to
have one, two or three independent variables, and all native or user-defined
fields known by \yt{} can serve as variables in this process.  Each profile,
defined uniquely by its bounds and independent variables, accepts a data
container as a source and is then self-contained within an instance of the
appropriate Python class (\texttt{BinnedProfile1D}, \texttt{BinnedProfile2D} or
\texttt{BinnedProfile3D}).

\begin{figure}
\begin{centering}
\includegraphics[width=0.48\textwidth]{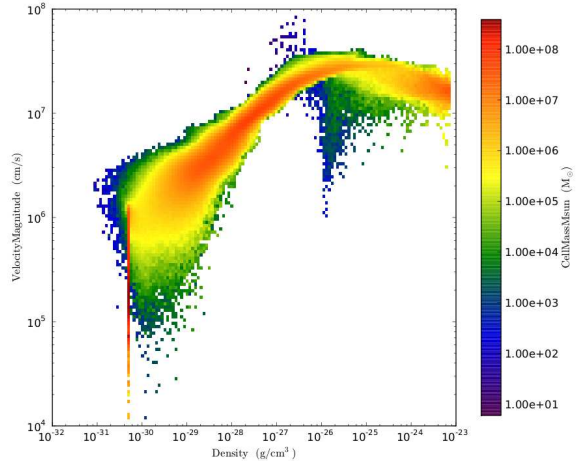}
\caption{An unweighted profile, showing the distribution of mass in a galaxy merger
simulation as a function of density (x-axis) and the velocity (y-axis).}
\label{fig:phase_mass}
\end{centering}
\end{figure}

One can imagine profiles serving two different purposes: to show the average
value of a variable at a fixed location in the phase space of a set of
independent variables (such as the average molecular hydrogen fraction as a
function of density and temperature), or for the sum of a variable inside a
phase space of independent variables (such as the total mass within bins of
density and temperature.) \yt{} can calculate both of these types of profiles
over any data container.  This process is essentially that of a weighted
histogram.  We define up to three axes of comparison, which will be designated
$x$, $y$, and $z$, but should not be confused with the spatial axes of the
simulation.  These are discretized into $x_0 ... x_{n-1}$ where $n$ is the
number of bins along the specified axis.  Indices $j$ for each value among the
set of points being profiled are then generated along each axis such that
\begin{equation}
  x_{j} \leq v_i < x_{j+1}.  
\end{equation}
These indices are then used to calculated the weighted average or sum in each
bin:
\begin{equation}
  V_j = \frac{\sum_{i=1}^{N} v_i w_i}{\sum_{i=1}^{N} w_i}
\end{equation}
where $V_j$ is now the average value in bin $j$ in our weighted average, and
the $N$ points are selected such that their index along the considered axis is
$j$.  This method is trivially extended to multiple dimensions.  To conduct a
non-averaged distribution, the weights are all set to $1.0$ in the numerator,
and the sum in the denominator is not calculated.  This allows, for example,
the examination of mass distribution in a plane defined by chemo-thermal
quantities.  In Figure~\ref{fig:phase_mass} we show an example image, where the
distribution of matter in a galaxy merger simulation has been shown as a
function of density and the local $v_{\mathrm{rms}} \equiv \sqrt{v_x^2 + v_y^2
+ v_z^2}$.  This is an ``unweighted'' profile, where the value in every cell
corresponds to the total mass occupying that region in phase space.

\subsection{Persistent Object Storage}

The construction of objects, particularly when guided by analysis pipelines or
calculated values, can often be a computationally expensive task; in
particular, clumps found by the contouring algorithm (see
Section~\ref{sec:level_sets}) and the gravitational binding checks that are
used to describe them require a relatively time-consuming set of steps.  To
save time and enable repeatable analysis, the storage of objects between
sessions is essential.  Python itself comes with an object serialization
protocol called ``pickle'' that can handle most objects.  However, by default
the pickle protocol is greedy -- it seeks to take all affiliated data.  For a
given \yt{} object, this may include the entire hierarchy of AMR grid data, the
parameter file describing the AMR run, all arrays associated with that object,
and even user-space variables (see \S~\ref{sec:codes} for more details about
the former two).  Under the assumption that the data used to generate the
fields within a given object will be available the next time the object is
accessed, we can reduce the size and scope of the pickling process by designing
a means of storing and retrieving these objects across sessions.

\yt{} implements a custom version of the pickling protocol, storing instead a
description in physical space of the object itself. This usually involves
replicating the parameters used to create the \texttt{AMRData} object, such as
the radius and center of a sphere.  Once the protocol has been executed, all
the information necessary to reconstruct the object is stored either in a
single, standalone file or in a local data store.

The primary obstacle to retrieving an object from storage is ensuring that the
affiliation of that object with a given dataset is retained.  This is
accomplished through a persistent per-user storage file, wherein unique IDs for
all ``known'' datasets are stored.  These unique IDs are updated when new
datasets are opened.  When an object is retrieved from storage, the unique ID
affiliated with that object is looked up and the dataset is opened and returned
to the user with the object.

\section{Visualization}\label{sec:visualization}

\yt{} provides methods for creating 2D and 3D visualizations of simulation
data.  The mechanisms for creating 2D visualizations have two primary
components: the data-handling portion and the figure creation or
``pixelization'' step.  The former is composed of a sophisticated set of
objects which provide uniform access to 2D data objects, while the latter is a
simple method for making plots quickly, which can be wrapped into other
convenience functions (both created by \yt{} and external to \yt{}.)  The
figure creation in \yt{} is motivated by a desire for simplicity: rather than
attempting to accommodate the myriad use cases and user preferences, \yt{}
seeks to provide a set of routines that can be extended easily.  Users
requiring complex figures for specific publications can take the 2D image pixel
buffers provided by \yt{} and feed them to any plotting package, though \yt{}
integrates most naturally with the Matplotlib Python module
\citep{matplotlib_paper}. Here, we first describe each of the 2D pixalization
mechanisms, and then the 3D volume rendering algorithms.  Futher information on
the simple, built-in figure generation can be found in the \yt{} documentation.

\subsection{Slices}

The simplest means of examining data is plotting grid-axis aligned slices
through the dataset.  This has several benefits - it is easy to calculate which
grids and which cells are required to be read off disk (and most data formats
allow for easy striding of data off disk, which reduces this operation's IO
overhead) and the process of stepping through a given dataset is relatively
easy to automate.

To construct a set of data points representing a slice, we construct a set of
data points defined as $(x_p, dx_p, y_p, dy_p, v)$ where $p$ indicates that
this is in the image plane rather than in the global coordiantes of the
simulation, and $v$ is the value of the field selected; furthermore, every
returned $(x_p,dx_p,y_p,dy_p,v)$ point does not overlap with any points where
$dx < dx_p$ or $dy < dy_p$.  Each point is at the finest resolution available.

\subsection{Projections}\label{sec:projections}

When handling astrophysical simulation data, one often wishes to examine either
the sum of values along a given sight-line or a weighted-average along a given
sight-line, in a \emph{projection}.  \yt{} provides an algorithm for generating
line integrals in an adaptive fashion, such that every returned
$(x_p,dx_p,y_p,dy_p,v)$ point does not contain data from any points where $dx <
dx_p$ or $dy < dy_p$; the alternative to this is a simple 2D image array of
fixed resolution perpendicular to the line of sight whose values are filled in
by all of the cells of the source object with overlapping domains.  But, by
providing this list of \emph{all} finest-resolution data points in a projected
domain, images of any field of view can be constructed essentially
instantaneously; conversely, however, the initial projection process takes
longer, for reasons described below.  We term the outputs of this  process
\emph{adaptive projections}.
For the Santa Fe Light Cone dataset \citep{2007ApJ...671...27H}, to project the
entire domain at the highest resolution would normally require an image with $2^30$
values.  Utilizing this adaptive projection method, we require less than $1\%$
of this amount of image storage.

To obtain the finest points available, the grids are iterated over in order of
the level of refinement -- first the coarsest and then proceeding to the finest
levels of refinement.  The process of projecting a grid varies depending on the
desired output from the projection.  For weighted averages, we first
evaluate the sums
\begin{equation}
\begin{array}{lcl}
V_{ij} & = & \sum_n v_{ijn}w_{ijn}dl \\
W_{ij} & = & \sum_n w_{ijn}dl
\end{array}
\end{equation}
where $V_{ij}$ is the output value at every cell in the image plane,
$W_{ij}$ is the sum of the weights along the line of sight in the
image plane, $v_{ijn}$
is every cell in the grid's data field, $w_{ijn}$ is the weight field at every
cell in the grid's data field, and $dl$ is the path length through a single
cell.  Because this process is conducted on a grid-by-grid basis, and the $dl$
does not change within a given grid, this term can be moved outside of the sum.
For an unweighted integration, $W_{ij}$ is set to $1.0$, rather than to the
evaluation of the sum.  A mask of cells where finer data exists is reduced with
a logical ``and'' operation along the axis of projection; any cell where this
mask is ``False'' has data of a higher refinement level available to it.  This
grid is then compared against all grids on the same level of refinement with
which it overlaps; the flattened $x$ and $y$ position arrays are compared via
integer indexing and any collisions are combined.  This process is repeated
with data from coarser grids that have been identified as having subsequent
data available to it; each coarse cell is then added to the $r^2$ cells on the
current level of processing, where $r$ is the refinement factor.  At this
point, all cells in the array of data for the current level where the reduced
child mask is ``True" are removed from subsequent processing, as they are part
of the final output of the projection.  All cells where the child mask is
``False" are retained to be processed on the next level.  In this manner, we
create a cascading refinement process, where only two levels of refinement have
to be compared at a given time.

When the entire data hierarchy has been processed, the final flattened arrays
of $V_p$ and $W_p$ are divided to construct the output data value $v$:
\begin{equation}
  v(x,y) = V(x,y)/W(x,y)  
\end{equation}
which is kept as the weighted average value along the axis of projection.  In
the case of an un-weighted projection, the denominator reduces to $\int dl$,
which is in fact unity.  Once this process is completed, the projection object
respects the same data protocol as an ordinary slice and can be plotted in the
same way.  Future versions of \yt{} will migrate to a quad-tree projection
mechanism, currently still in the testing phase.  Using this quad-tree approach
will allow grids to be handled in any order, as well as providing an overall
speed increase.

\subsection{Image Creation}\label{sec:image_creation}

Because of the multi-scale nature of most adaptive mesh refinement
calculations, \yt{} operates in a manner to reduce the total disk access
required to generate 2D visualizations.  Pragmatically, this means that slices
and projections are constructed of the finest available data at all points and
then a pixel buffer is created for display.  This enables the user to conduct a
single projection through a dataset and then, with minimal computation effort,
create many images of different regions in that domain.  For central-collapse
simulations, a single slice can be made through the data and then images can be
made of that slice at different widths without handling the 3D data again.
In \yt{}, 2D data sources are stored as a flattened array of data points, with
the positions and widths of those data points.  To construct an image buffer,
these data points are pixelized and placed into a fixed-resolution array,
defined by $(x_{p,\mathrm{min}},x_{p,\mathrm{max}}, y_{p,\mathrm{min}},
y_{p,\mathrm{max}})$.  Every pixel in the image plane is iterated over, and any
cells that overlap with it are deposited into every pixel $I_{ij}$ as

\begin{eqnarray}
\alpha & = & A_c / A_p \\
\alpha v & \rightarrow & I_{ij}
\end{eqnarray}

where $\alpha$ is an attempt to anti-alias the output image plane, to account
for misalignment in the image and world coordinate systems and $A_c$ and $A_p$
are the areas of the cell and pixel respectively.  This process is generally
quite fast, and even for very large simulations (such as in
\citep{2007ApJ...671...27H}) the process of generating a pixel buffer from an
adaptive projection takes much less than one second.  The buffer created by
this process can be used either in \yt{}, utilizing the built-in methods for
visualization, or it can be exported to external utilities such as DS9
\citep{2003ASPC..295..489J}.

\subsection{Cutting Planes}

At some length scales in star formation problems, gas is likely to collapse
into a disk, which is often not aligned with the cardinal axes of the
simulation.  By slicing along the axes, patterns such as spiral density waves
could be missed and remain unexamined.  In order to better visualize off-axis
phenomena, \yt{} is able to create images misaligned with the axes.  A
\emph{cutting plane} is an arbitrarily-aligned plane that transforms the
intersected points into a new coordinate system such that they can be pixelized
and made into a publication-quality plot.

\begin{figure}
\begin{centering}
\includegraphics[width=0.48\textwidth]{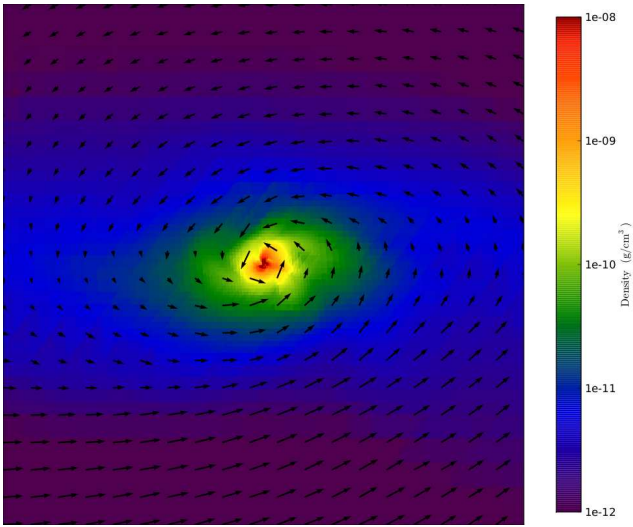}
\caption{An example oblique slice through a primordial star forming region, where the
image plane has been chosen such that its normal is coincident with the angular
momentum vector.  Velocity vectors have been overlaid.}
\label{fig:cutting}
\end{centering}
\end{figure}

To construct a \emph{cutting plane}, both a central point and a single normal
vector are required.  The normal vector is taken as orthogonal to the desired
image plane.  This leaves a degree of freedom for rotation of the image plane
about the normal vector and through the central point.  A minimization
procedure is conducted to determine the appropriate ``North" vector in the
image plane; the axis with which the normal vector ($\mathbf{n}$)
has the greatest cross product is selected and referred to as $\mathbf{a_0}$.
In addition to this, we define
\begin{equation}
  \begin{array}{lclcl}
    \mathbf{p_x} & = &  \mathbf{a_0} & \mathbf{\times} & \mathbf{n} \\
    \mathbf{p_y} & = &  \mathbf{n}   & \mathbf{\times} & \mathbf{p_x} \\
    \mathbf{d}   & = & -\mathbf{c}   & \mathbf{\cdot}  & \mathbf{n}
  \end{array}  
\end{equation}
where $\mathbf{c}$ is the vector to the center point of the plane, and
$\mathbf{d}$ is the inclination vector.  From this we construct two matrices,
the rotation matrix:
\begin{equation}
R = \left(\begin{array}{ccc}
p_{xi} & p_{xj} & p_{xk} \\
p_{yi} & p_{yj} & p_{yk} \\
n_{i} & n_{j} & n_{k}
\end{array}\right)  
\end{equation}
and its inverse, which are used to rotate coordiantes into and out of the image
plane, respectively.  Grids are identified as being intersected by the cutting plane
through fast array operations on their boundaries.  We define a new array, $D$, where
\begin{equation}
D_{ij} = \mathbf{v}_{ji} \mathbf{\cdot} \mathbf{d}  
\end{equation}
where the index $i$ is over each grid and the index $j$ refers to which of the eight
grid vertices ($\mathbf{v}$) of the grid is being examined.  Grids are accepted if
all three components of every $D_{j}$ are of identical sign:
\begin{equation}
  \mathrm{all}( D_j < 0 ) \mathrm{or~all} (D_j > 0).
\end{equation}
Upon identification of the grids that are intersected by the cutting plane, we select
data points where
\begin{equation}
  ||\mathbf{p} \mathbf{\cdot} \mathbf{n} + \mathbf{d} || <
  \frac{\sqrt{dx^2+dy^2+dz^2}}{2}.
\end{equation}
This generates a small number of false positives (from regarding a cell as a sphere
rather than a rectangular prism), which are removed during the pixelization step when
creating a plot.  Each data point is then rotated into the image plane via the
rotation matrix:
\begin{equation}
  \begin{array}{lcl}
    \mathbf{p} \mathbf{\cdot} \mathbf{p}_x & \rightarrow & x_p \\
    \mathbf{p} \mathbf{\cdot} \mathbf{p}_y & \rightarrow & y_p.
  \end{array}
\end{equation}
This technique requires a new pixelization routine in order to ensure that the
correct cells are taken and placed on the plot, which requires an additional
set of checks to determine if the cell intersects with the image plane.  The
process here is similar to the standard pixelization procedure, described in
Section~\ref{sec:image_creation}, with the addition of the rotation step.
Defining $d = \sqrt{dx^2+dy^2+dz^2}$, every data point where $(x_p \pm d,y_p
\pm d)$ is within the bounds of the image is examined by the pixelization
routine for overlap of the data point with a pixel in the output buffer.  Every
potentially intersecting pixel is then iterated over and the coordinates $(x_i,
y_i, 0)$ of the image buffer are rotated via the inverse rotation matrix back
to the world coordinates $(x', y', z')$.  These are then compared against the
$(x, y, z)$ of this original datapoint.  If all three conditions 
\begin{equation}
  \begin{array}{lcl}
    |x-x'| & < & dx \\
    |y-y'| & < & dy \\
    |z-z'| & < & dz
  \end{array}
\end{equation}
are satisfied, the data value from the cell is deposited in that image buffer
pixel.  An unfortunate side effect of the relatively complicated pixelization
procedure, as well as the strict intersection-based inclusion, is that the
process of antialising is non-trivial and computationally expensive, and
therefore \yt{} does not perform any antialiasing of cutting-plane images.  By
utilizing the same transformation and pixelization process, overlaying in-plane
velocity vectors is trivially accomplished and a simple mechanism to do so is
included in \yt{}.  In Figure~\ref{fig:cutting} we show an example image, where
the inner $100~\mathrm{AU}$ of a primordial star forming region is shown, where
the normal to the image plane is aligned with the angular momentum vector and
where velocity vectors have been overlaid.

\subsection{Volume Rendering}\label{sec:volume_rendering}

Direct ray casting through a volume enables the generation of new types of
visualizations and images describing a simulation.  \yt{} has the facility to
generate volume renderings by a direct ray casting method.  Currently the
implementation is implemented to run exclusively on the CPU, rather than faster
hardware-based rendering mechanisms, but this also allows for clearer
descriptions of the algorithms used for compositing, calculation of the
transfer function, and future advances in parallelization.  Furthermore, it
eases the task of informing volume renderings with other analysis results: for
instance, halo location, angular momentum, spectral energy distributions and
other derived or calculated information.

The volume rendering in \yt{} follows a relatively straightforward approach.

\begin{enumerate}
\item Create a set of transfer functions governing the emission and absorption
      in red, green, blue, $\alpha$ space (rgba) as a function of one or more
      variables. ($f(v) \rightarrow (r,g,b,a)$) These can be functions of any field
      variable, weighted by independent fields, and even weighted by other evaluated
      transfer functions.
\item Partition all grids into non-overlapping, fully domain-tiling ``bricks."
      Each of these ``bricks" contains the finest available data at any
      location.  This process itself is the most time consuming, and is
      referred to as a process of homogenization.
\item Generate vertex-centered data for all grids in the volume rendered domain.
\item Order the bricks from back-to-front.
\item Construct plane of rays parallel to the image plane, with initial values set
      to zero and located at the back of the region to be rendered.
\item For every brick, identify which rays intersect.  These are then each `cast'
      through the brick.

  \begin{enumerate}
  \item Every cell a ray intersects is sampled 5 times (adjustable by parameter),
        and data values at each sampling point are trilinearly interpolated from
        the vertex-centered data. This is needed when the transfer
        function is composed of very thin contours which might not be
        picked up by a single sample of the cell.
  \item Each transfer function is evaluated at each sample point.  This gives us,
        for each channel, both emission ($j$) and absorption
        ($\alpha$) values.
  \item The value for the pixel corresponding to the current ray is updated with
        new values calculated by rectangular integration over the path length:

        $$v^{n+1}_{i} =  j_{i}\Delta t + (1 - \alpha_{i}\Delta t )v^{n}_{i}$$

        where $n$ and $n+1$ represent the pixel before and after
        passing through a sample, $i$ is the color (red, green, blue) and 
        $\Delta t$ is the path length between samples.
  \end{enumerate}
\end{enumerate}

At this point, the final resultant plane of values is returned to the user.
Because this process is neutral to the colors used, and because it integrates a
simplified form of the radiative transfer equation, it can be repurposed for
generating simulated images from realistic input emissions and absorptions, in
the absence of scattering terms.

\begin{figure}
\begin{centering}
\includegraphics[width=0.48\textwidth]{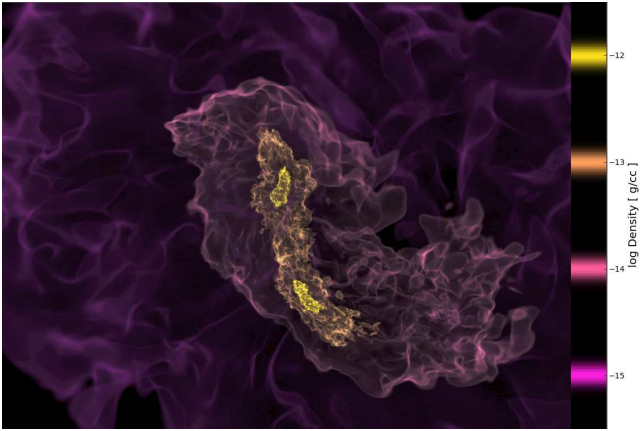}
\caption{A volume rendering of a metal-free star forming region that has fragmented
into two cores, each of which is likely to host a Population III star.  The field of
view is $2000~\mathrm{AU}$.  Isocontours were placed at $10^{-15}, 10^{-14},
10^{-13}$ and $10^{-12}~\mathrm{g}~\mathrm{cm}^{-3}$ \citep{2009Sci...325..601T}.}
\label{fig:vr_binary}
\end{centering}
\end{figure}

In \yt{}, volume rendering is exposed through both a simplified interface,
wherein images are generated and returned.  A more detailed ``Camera''
interface that allows for camera paths, zooms, stereoscopic rendering and
easier access to the underlying vector plane is also available.  Transfer
functions that can automatically sample colormaps as well as one that provides
off-axis line integrals are supplied, as well as a transfer function whose
colors correspond to Johnson filter-convolved Planck emission with approximate
scattering terms, as in \cite{vg06-kaehler}.

By allowing for detailed control over the specification of the transfer
function, viewing angle and generation of images, volume renderings that
contain a scientific narrative are easier to create.  For instance, in
Figure~\ref{fig:vr_binary} we have constructed a volume rendering of the
Population III star formation simulation described in
\cite{2009Sci...325..601T}, where a collapsing metal-free halo has been found
to fragment into two distinct clumps.  This volume rendering has been aligned
such that the normal vector to the image plane is aligned with the angular
momentum vector of the two-clump system.  Furthermore, the isocontours visible
in the image have been selected such that they coincide with transitions
between chemical states in the cloud.  Additional volume renderings based on
derived fields describing chemical and kinetic quantities could be constructed,
as well.

\section{Simulation Codes}\label{sec:codes}

\yt{} was originally designed exclusively for analyzing data from the adaptive
mesh refinement code Enzo.  The built-in fields were tuned to the needs of Enzo
analysis and the disk IO mechanisms were tuned for Enzo data formats.  However,
abstractions to the underlying representation of state variables have enabled
it to be extended to work with a number of other simulation codes natively,
including Orion \citep{1998ApJ...495..821T, 1999ASSL..240..131K,
2004ApJ...611..399K, 2007ApJ...667..626K}, FLASH \citep{2000ApJS..131..273F},
Chombo \citep{chombo_website} and RAMSES \citep{2002A&A...385..337T}.  Work is
ongoing to add support for the ART \citep{1997ApJS..111...73K} and Gadget
\citep{2005MNRAS.364.1105S} simulation codes.  This cross-code support has
already enabled collaboration between these communities, and it is our hope
that it will act as a gateway to better development of common infrastructure
for analysis of simulation outputs. 

\subsection{Code Support}

For a simulation code to be considered supported, the following
requirements must be met:

\begin{itemize}
\item Simulation-specific fluid quantities and domain information must be
translated or mapped into the common yt format.  This enables abstraction of
the underlying quantities in a common (typically cgs) system.
\item Data on disk must be mapped and localized in memory; this enables \yt{} to
find and read data from the disk in order to present it to the user.  For AMR
codes, this means identifying grid patches or blocks and localizing them to a
region on disk.
\item Routines for reading data must be constructed, to read either entire grid
patches or subsets of those grid patches.
\end{itemize}

Adding support for a new code requires the implementation of a set of
subclasses that govern the interface between the data on disk and the internal
\yt{} structures; in the \yt{} distribution this is documented and examples are
provided.  These data structures and IO routines are compartmentalized in the
\yt{} source code.

By abstracting these three functions into code-specific routines, and ensuring
a uniform set of units and fluid quantity meanings, the interface to data (and
thus the analysis of the data) becomes agnostic to the simulation code.  The
same routines for calculating the moment of inertia of a protostellar accretion
disk in an Orion simulation can be utilized to calculate the moment of inertia
of a protostellar accretion disk in an Enzo simulation.  The routines for
calculating the clumping factor in a RAMSES simulation can be used on a FLASH
simulation.  Cross-code comparison of results is possible with minimal effort:
not only do indepdendent research groups no longer have to reinvent identical
routines for executing common analysis tasks, but they can be certain the
specific details of implementations of these routines are identical.

RAMSES is based on an octree data structure.  In order to support the RAMSES
code (and other octree codes), the \yt{} backend reads the hierarchy of cells
and then conducts a process of patch coalescing.  To identify patches, we have
implemented the mechanism used by Enzo to create subgrid patches from cells
flagged for refinement.  In this algorithm, one-dimensional histograms of cell
counts are first calculated along all three dimensions.  These histograms are
inspected for zeros and then for the strongest zero-crossings in their second
derivative.  At these locations, cutting planes are inserted.  This process is
conducted recursively until the ratio of finest cells in a region to the total
number of cells in that region is greater than some efficiency factor,
typically set to be 0.2.  These patches are then used as the final grid
patches, allowing array operations on the enclosed cells to operate in bulk.

\subsection{Grid Data Format}

In order to enable analysis of the broadest possible number of simulation
codes, \yt{} supports the reading of a generic gridded data format, based on
HDF5\footnote{\url{http://www.hdfgroup.org/}}, described in the \yt{}
documentation.  This format explicitly notes conversion factors, field
positions, particle positions and has mechanisms for explicitly adding new
fields, where the field is named, units are described and it is explicitly
included.  \yt{} is able to write other formats to this highly-explicit format,
enabling conversion between different simulation formats.

Furthermore, because this is a fully-explicit format, external codes for which
native support in \yt{} is not available can be converted to this intermediate
format, where they can then be read in and analyzed in \yt{}.  

\subsection{Particle Data}

Even grid-based codes, wherein fluid quantities are repsented in an Eulerian
fashion with the fluid quantities defined in a mesh everywhere in the domain,
are typically in fact hybrid codes combining both particles and mesh quantities
in a single set of coupled governing equations.  Particles are used represent
the collisionless (dark matter, stellar) components of the calculation, while
the fluid quantities represent the gaseous component.  To accommodate this,
\yt{} can read particle values as well as gridded fluid values off disk.  When
a particle field (for example mass, position, velocity, age of star particles,
luminosity) is requested from a data container, \yt{} identifies those
particles that reside within the data container, loads the requested field from
disk, and returns this to the user.  \yt{} additionally provides particle
interpolation and cloud-in-cell deposition routines, so that particles can be
deposited onto grid patches and their attributes regarded as fluid, rather than
discrete, quantities.

With data created by Enzo and FLASH, particles are associated on disk within
the highest-resolution grid patch in which they reside; this allows \yt{} to
conduct fast, on-demand loading of dark matter and star particles.  For data
created by Orion, sink and star particles are stored in a separate, unique
file, which can be read into memory and associated with data containers as
necessary.  For Enzo data, several data container-aware routines are provided
to enable very fast intra-grid selection of particles within data containers
such as spheres and rectangular prisms.  However, while load-balancing for
fluid fields can be estimated in advance, load-balancing of particle analysis
requires more care.  These challenges are discussed in more detail in
\citet{2010arXiv1001.3411S}.

Particle fields can also be used as input into derived field creation.  For
instance, many star formation prescriptions in cosmological codes
\citep[e.g.,][]{1992ApJ...399L.113C} rely on an initial mass at the time of
creation of a star particle, which is then dwindled over time as the star
particle feeds material back into its environment.  The initial mass is
therefore encoded in the combination of the age and the creation time of a star
particle, and a derived field can be constructed specific to the star formation
prescription to provide the initial mass of the star particles.  By combining
derived fields for spectral energy distributions with the particle deposition
routines provided in \yt{}, star particles can also be used as an input to the
volume rendering algorithm (\S~\ref{sec:volume_rendering}).

\section{Parallelism}\label{sec:parallelism}

As the capabilities of supercomputers grow, the size of datasets grows as well.
Most standalone codes are not parallelized; the process is time-consuming,
complicated, and error-prone.  Therefore, the disconnect between simulation
time and data analysis time has grown ever larger.  In order to meet these
changing needs, \yt{} has been modified to run in parallel on multiple
independent processing units on a single dataset.  Specifically, utilizing the
Message Passing Interface \citep[][hereafter MPI]{MPIStandard} via the
\texttt{mpi4py} Python module
\citep[][\url{http://mpi4py.googlecode.com/}]{MPI4PY:PAPER1, MPI4PY:PAPER2},
a lightweight, NumPy-native wrapper that enables natural access to the C-based
routines for interprocess communication, the code has been able to subdivide
datasets into multiple decomposed regions that can then be analyzed
independently and joined to provide a final result.  A primary goal of this
process has been to preserve at all times the API, such that the user can
submit an unchanged serial script to a batch processing queue, and the toolkit
will recognize it is being run in parallel and distribute tasks appropriately.

The tasks in \yt{} that require parallel analysis can be divided into two broad
categories: those tasks that act on data in an unordered, uncorrelated fashion
(such as weighted histograms, summations, and some bulk property calculation),
and those tasks that act on a decomposed domain (such as halo finding and
projection).  All objects and tasks that utilize parallel analysis exist as
subclasses of \texttt{ParallelAnalysisInterface}, which provides a number of
functions for load balancing, inter-process communication, domain decomposition
and parallel debugging.  Furthermore, \yt{} itself provides a very simple
parallel debugger based on the Python built-in \texttt{pdb} module.

\subsection{Unordered Analysis}\label{sec:unordered_analysis}

To parallelize unordered analysis tasks, a set of convenience functions have
been implemented utilizing an initialize/finalize formalism; this abstracts the
entirety of the analysis task as a transaction.  Signaling the beginning and
end of the analysis transaction initiates several procedures, defined by the
analysis task itself, that handle the initialization of data objects and
variables and that combine information across processors.  These are abstracted
by an underlying parallelism library, which implements several different
methods useful for parallel analysis.  By this means, the intrusion of
parallel methods and algorithms into previously serial tasks is kept to a
minimum; invasive changes are typically not necessary to parallelize a task.
This transaction follows several steps:
\begin{enumerate}
\item Obtain list of grids to process
\item Initialize parallelism on the object
\item Processes each grid
\item Finalize parallelism on the object
\end{enumerate}
This is implemented through the Python iterator protocol; the initialization of
the iterator encompasses the first two steps and the finalization of the
iterator encompasses the final step.

Inside the grid selection routine, \yt{} decomposes the relevant set of grids
into chunks based on the organization of the datasets on disk.  Implementation
of the parallel analysis interface mandates that objects implement two
gatekeeper functions for both initialization and finalization of the parallel
process.  At the end of the finalization step, the object is expected to be
identical on all processors.  This enables scripts to be run identically in
parallel and in serial.  For unordered analysis, this process results in
close-to-ideal scaling with the number of processors.

In order to decompose a task across processors, a means of assigning grids to
processors is required.  For spatially oriented-tasks (such as projections)
this is simple and accomplished through the decomposition of some spatial
domain.  For unordered analysis tasks, the clear means by which grids can be
selected is through a minimization of file input overhead.  The process of
reading a single set of grid data from disk can be outlined as:

\begin{enumerate}
\item Open file
\item Seek to grid data position
\item Read data
\item Close file
\end{enumerate}

For those data formats where multiple grids are written to a single file, this
process can be consolidated substantially by performing multiple reads inside a
single file once it has been opened.  If we know the means by which the grids
and fields are ordered on disk, we can simplify the seeking requirements and
instead read in large sweeps across the disk.  By futher pre-allocating all
necessary memory, this becomes a single operation that can be accomplished in
one ``sweep'' across each file.  By allocating as many grids from a single
``grid output'' file on a single processor, this procedure can be used to
minimize file overhead on each processor.  Each of these techniques
are implemented where possible. 

\begin{figure}
\begin{centering}
\plotone{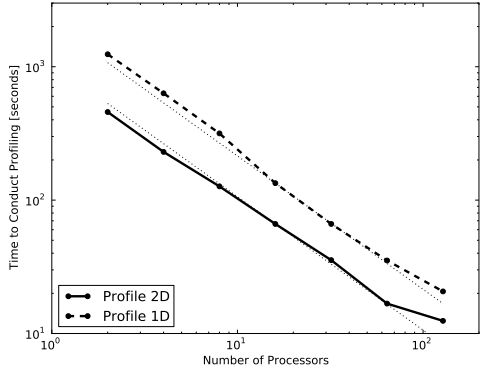}
\caption{Time taken for conducting 1- and 2-D profiles on the Santa Fe Light
Cone dataset at $z=0$ \citep{2007ApJ...671...27H}, a $512^3$ dataset with 6
levels of refinement (throughout the entire simulation domain) and a total of
$5.5\times10^8$ computational elements.  The overplotted thin solid lines
represent ideal scaling, as calibrated to the time taken by 16 processors.}
\label{fig:scaling_profile}
\end{centering}
\end{figure}

In Figure~\ref{fig:scaling_profile} we show the results of a strong-scaling
study of conducting profiles of the final dataset from the Santa Fe Light Cone
\citep{2007ApJ...671...27H} project.  This dataset consists of
$5.5\times10^{8}$ computational elements.  The dashed black corresponds to
profiling in one dimension, and the solid line corresponds to profiling in two
dimensions.  Overplotted in thin solid lines are the ideal scaling
relationships, as calibrated to the time taken by 16 processors.  We see nearly
ideal strong scaling up to 128 processors, at which point overhead dominates;
we are essentially starving the processors of work at this scale.  The overall
time taken to conduct a profile is quite low, on one of the largest AMR
datasets in the published literature.  We note also that the substantial speed
difference between the two mechanisms of profiling, which is counter-intuitive,
is a result of a difference in implementation of the histogramming method; 1D
profiles use a pure-python solution to histogramming, whereas 2D profiles use a
hand-coded C routine for histogramming.  Future versions of \yt{} will
eliminate this bottleneck for 1D profiling and we expect to regain parity
between the two methods.

\subsection{Spatial Decomposition}\label{sec:spatial_decomposition}

Several tasks in \yt{} are inherently spatial in nature, and thus must be
decomposed in a spatially-aware fashion.  MPI provides a means of decomposing
an arbitrary region across a given number of processors.  Through this method,
the \texttt{ParallelAnalysisInterface} provides mechanisms by which the domain
can be divided into an arbitrary number of subdomains, which are then realized
as individual data containers and independently processed.

For instance, because of the inherently spatial nature of the adaptive
projection algorithm implemented in \yt{}, parallelization requires
decomposition with respect to the image plane (however, as noted in Section
\ref{sec:projections} future revisions of the algorithm will allow for
unordered grid projection.)  To project in parallel, the computational domain
is divided such that the image plane is distributed equally among the
processors; each component of the image plane is decomposed into rectangular
prisms (\texttt{AMRRegion} instances) along the entire line of sight.  Each
processor is allocated a rectangular prism of dimensions $(L_i, L_j, L_d)$
where the axes have been rotated such that the line of sight of the projection
is the third dimension, $L_i \times L_j$ is constant across processors, and $L_d$ is
the entire computational domain along the axis of projection.  Following the
projection algorithm, each processor will then have a final image plane set of
points, as per usual: $$ (x_p, dx_p, y_p, dy_p, v) $$ but subject to the
constraints that all points are contained within the rectangular prism as
prescribed by the image plane decomposition.  At the end of the projection step
all processors join their image arrays, which are guaranteed to contain only
unique points.

Enzo and Orion utilize different file formats, but both are designed to output
a single file per processor with all constituent grids computed on that
processor localized to that file.  Both codes conduct ``load balancing"
operations on the computational domain, so processors are not necessarily
guaranteed to have spatially localized grids; this results in the output format
not being spatially decomposed, but rather unordered.  As a result, this method
of projection does not scale as well as desired, because each processor is
likely to have to read grid datasets from many files.  Despite that, the
communication overhead is essentially irrelevant, because the processors only
need to communicate the end of the projection process, to share their
non-overlapping final result with all other processors in the computational
group.

\begin{figure}
\begin{centering}
\plotone{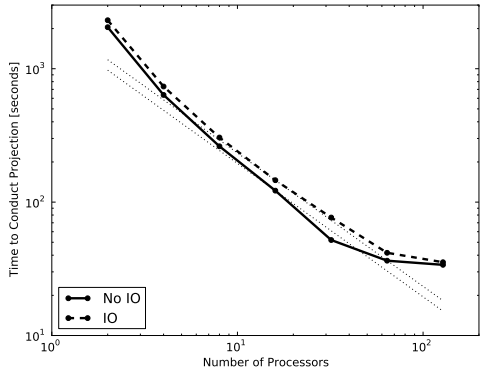}
\caption{Time taken creating adaptive projections of the Santa Fe Light Cone
dataset at $z=0$ \citep{2007ApJ...671...27H}, a $512^3$ dataset with 6 levels
of refinement (throughout the entire simulation domain) and a total of
$5.5\times10^8$ computational elements.  In the case where IO was not
conducted, a field consisting uniformly of 1.0 everywhere was used as input.
The overplotted thin lines represent ideal scaling, as calibrated to the time
taken by 16 processors.}
\label{fig:scaling_projection}
\end{centering}
\end{figure}

In Figure~\ref{fig:scaling_projection} we show the results of a strong-scaling
study of adaptively projecting the same dataset as in
Section\ref{sec:unordered_analysis}.  The dashed line represents a projection of
the density field, whereas the solid line represents projection in the absence of
disk IO.  Clearly the algorithmic overhead dominates the cost of disk IO, but
we also see strong scaling between 4 and 64 processors; at 128 processors we
see deviation from this.  The relatively early termination of strong scaling
(64 processors for this dataset, but we expect this to be higher for larger
datasets) as a result of algorithmic overhead is one of the motivations behind
future improvements to the projection algorithm, as discussed in
Section~\ref{sec:projections}.  However, from a pragmatic perspective, because
\yt{} creates adaptive projections, the time taken to project is a one-time
investment and thus not a rate-determining step for post-processed
analysis.  For non-adaptive projections, the process of handling all of the
data, conducting parallel reductions and outputting images must be undertaken
for every chosen field of view.

\section{Simulation Code Embedding}\label{sec:embedding}

An outstanding problem in the analysis of large scale data is that of
interfacing with disk storage; while data can be written to disk, read back,
and then analyzed in an arbitrary fashion, this process is not only slow but
requires substantial intermediate disk space for a substantial quantity of data
that will undergo severely reductionist analysis \citep{2007arXiv0705.1556N}.
To address this problem, the typical solution is to insert analysis code,
generation of derived quantities, images, and so forth, into the simulation
code.  However, the usual means of doing this is through either a substantial
hand-written framework that attempts to account for every analysis task, or a
limited framework that only handles very limited analysis tasks.  \yt{}
provides an explicit embedding API to enable in-line analysis.

By enabling in-line analysis, the relative quantity of analysis output is
substantially greater than that enabled by disk-mediated analysis; the cadence
of analysis tasks can be increased, leading to greater time-domain resolution.
Removing numerous large files dumped to disk as a prerequisite for conducting
analysis and generating visualization allows for a much more favorable ratio of
data to analyzed data.  For example, in a typical Population III star formation
simulation, such as in \citet{2009Sci...325..601T}, the size of the data dumps
can be as much as 10 gigabytes per timestep; however, the relative amount of
information that can be gleaned from these outputs is significantly smaller
\citep{2009Sci...325..601T}.  Using smaller data output mechanisms as well as
more clever streaming methods can improve this ratio; however, by enabling
in-line analysis, images of the evolution of a collapsing Population III halo
can be output at every single update of the hydrodynamical time, allowing for
true ``movies'' of star formation to be produced.  By allowing for the creation
and exporting of radial profiles and other analytical methods, this technique
opens up vast avenues for analysis while simulations are being conducted,
rather than afterward.

The Python/C API allows for passage of data in-memory to an instance of the
Python interpreter.  \yt{} has been instrumented such that it can be accessed
by an embedded Python interpreter inside a simulation code, such that one
interpreter instance exists for every MPI task.  \yt{} provides a clear API for
passing the necessary geometric information from the simulation code to the
analysis package.  By utilizing thin wrappers around the memory in which field
values and simulation data exist, the contents of the running simulation are
exposed to \yt{} and analysis can be conducted on them.  While this currently
works for many relatively simple tasks, it is not currently able to decompose
data spatially; as we are constrained by the parallel nature of most domain
decomposition algorithms, we attempt to avoid passing data between MPI tasks.
This means if a grid resides within MPI task 1, it will not be passed to MPI
task 2 during the analysis stage.  Currently this mechanism for inline analysis
has been exposed to Enzo simulations, and we hope to extend this in the future
to additional simulation codes.

Inline analysis will only become more important as simulations increase in size
and scope, and future development in \yt{} will make it easier, more robust,
and more memory efficient.  The primary mechanism by which \yt{} will be
embedded will change; future iterations of the inline analysis interface will
rely on communication between separate MPI jobs for simulation and analysis,
rather than an analysis task that shares memory space with the running
simulation code.  This mechanism will allow asynchronous analysis tasks to
be run, enabling the simulation to proceed while the user controls the data
that is examined.  Additionally, the method for interfacing \yt{} and
simulation codes will be provided as a single C++ library that can be
linked against, allowing it to be embedded by other developers.

\section{Description of Selected Analysis Modules}\label{sec:analysis_modules}

As a result of the ability to assemble complicated chains of analysis tasks
from component parts, \yt{} has accumulated a number of pre-defined analysis
modules.  These modules are included with the base distribution of \yt{} and
are designed to provide a number of entry points and in some cases even
interact with each other.  Adding a new analysis module is a straightforward
process; the specifics of the application programming interface (e.g., the
required paramters and returned values) must be documented and made available
in a public-facing function, with appropriate documentation.  The code for all
analysis modules is required to export this API, which is then made available to
users.

Below, we describe a selection of the most mature and broadly-useful of
the analysis modules provided in the primary \yt{} distribution.  For a more
complete listing, we direct interested readers to the online \yt{}
documentation.

\subsection{Halo Finding}\label{sec:halo_finding}

In cosmological hydrodynamic simulations, dark matter particles and gas parcels
are coupled through gravitational interaction.  Furthermore, dark matter
dominates gravitational interaction on all but the smallest scales. Dark matter
particles act as a collisionless fluid, and are the first component of the
simulation to collapse into identifiable structures; as such, they can be used
effectively to identify regions of structure formation.  

The HOP algorithm \citep{eishut98} is an effective and tested means of
identifying collapsed dark matter halos in a simulation, and has been a part of
the Enzo code distribution for some time.  Typically to conduct halo finding, a
simulation is allowed to execute to completion, an entire dataset is loaded
into memory, and then the HOP algorithm processes the entire domain.  This
process is memory-intensive, and requires that the entire dataset be loaded
into a single computer.  It is not inherently parallel and thus does no domain
decomposition.  Including this code inside \yt{}, as a means of abstracting
away compilation and data access, was trivial; however, to do so the input to
HOP was generalized to be an arbitrary three-dimensional data source.  As a
result, the HOP algorithm can now be applied on subsets of the domain.  

Each identified halo (\texttt{Halo} objects) is a fully-qualified data source,
which can be used throughout the rest of \yt{}.  For instance, halos can be
identified by examining particle distributions, and then the constituent gas
quantities can be examined or visualized.

\yt{} provides standard HOP and Friends-of-friends algorithms, as well  a
ground-up parallel reimplementation of the HOP algorithm designed to be run on
very large datasets.  For a deeper discussion of this implementation, see
\cite{2010arXiv1001.3411S}.

\subsection{Halo Analysis}

Further analysis of halos can be performed in an automated way using \yt{}'s
halo profiling tool.  The halo profiler reads in a list of halos created by any
halo finding procedure.  The halo profiler may also be configured to run any of
\yt{}'s halo finders if halo information does not yet exist.  One dimensional
radial profiles and projections of user-specified fields are then made for each
halo in the list.  Because halos are typically quite small in relation to the
total computational domain, the halo profiler runs in parallel by distributing
the individual halos over the available processors.

A single dataset may contain thousands or tens of thousand of halos, in which
case insightful analysis often relies on the ability to extract scalar
quantities from each halo, such as the virial mass and radius or parameter
values for analytical models.  To facilitate this, simple filtering function
can be created whose only requirement is to accept a one dimensional profile
object.  These filter functions return True or False to indicate whether the
halo meets certain criteria and may optionally also return a dictionary of
scalar quantities.  An unlimited number of filters can be applied during the
profiling process.  When profiling has completed, a list of halos that passed
all of the filters is written out, including any quantities return by the
filter functions.  Below is an example of the output from the profiling and
filtering process.  In this example, a filter is used to calculate virial
quantities by interpolating profile data (here radius, stellar mass, and total
mass) at the point where the halo overdensity (also a profile field) crosses a
critical value.  This filter is also configured to accept only halos with total
virial mass greater than $10^{14} M_{\odot}$.  Note that some halos have been
rejected by the filter.

{\tiny\begin{verbatim}
# id  center[0]  center[1]  center[2]  RadiusMpc  StarMassMsun  TotalMassMsn
0000  0.706602   0.485794   0.466725   2.838402   2.628117e+13  1.052590e+15
0002  0.939776   0.664665   0.831547   2.461491   1.962899e+13  6.714788e+14
0004  0.809778   0.101728   0.708202   2.224953   1.712548e+13  5.095553e+14
0006  0.510057   0.268133   0.057933   2.286010   1.412319e+13  5.400834e+14
0007  0.205331   0.064149   0.764243   2.169436   1.212237e+13  4.662484e+14
\end{verbatim}}

\subsection{Merger Trees}

The dark and baryonic mass of a halo is accumulated in two non-exclusive ways.
Matter may simply be accreted onto the halo from the surrounding medium, or
multiple halos may merge together over time to form a larger single halo.
Because the mass accretion history drives the observable properties of the
galaxy embedded in the dark matter halo, it is important to be able to track
when, and by what means, mass is added to a halo.  The merger tree toolkit in
\yt{} enables the creation, analysis, and simple visualization of a merger tree
for the halos in a cosmological simulation.

The creation of the merger tree is fully-parallelized, and will automatically
call any of the parallelized \yt{} halo finders if the halos are not
pre-identified.  The output is a SQLite\footnote{\url{http://sqlite.org/}}
database, which provides a convenient and powerful way to store the halo
relational data.  SQL databases are a common way to store data of this type.
For example, it is the way that the Millennium Simulation
\citep{2005Natur.435..629S} merger tree data is
distributed\footnote{\url{http://galaxy-catalogue.dur.ac.uk:8080/Millennium/}}.

Included in the toolkit are several convenience functions that assist the user
in extracting data from the database.  The merger tree can drive powerful
analysis pipelines which use the many toolkits in \yt{} to analyze the dark and
baryonic matter content of the halos.  There is also a function that will
output a graphical representation of the merger tree in the
Graphviz\footnote{\url{http://graphviz.org/}} directed graph visualization
format.

\subsection{Two Point Functions}

A two point function operates on the field values at a pair of points separated
by some distance.  Examples include two point correlations of galaxies or
structure functions, such as the RMS gas velocity structure function used to
study the cascade of turbulent energy \citep{2007ApJ...665..416K}.  The two
point function toolkit in \yt{} is a framework that supports an unlimited
number of user-defined functions, and is fully-parallelized.

Conceptually, the two point function toolkit is simply a mechanical base upon
which a user may place any number of functions for evaluation.  Following a
defined functional input and output stencil, the user needs only to write
functions for their analyses.  The toolkit handles the data input, output, and
parallelism without direct involvement of the user.  A two point function that
runs on a small dataset on a personal computer will work on a massive dataset
in parallel on a supercomputer without any modifications.

Both unigrid and AMR datasets are supported, automatically and transparently to
the user.  The toolkit is highly adaptable and configurable.  How many times
and over what physical range the functions are evaluated is controlled by the
user.  The domain decomposition and level of parallelism is adjustable in an
intuitive and simple way.

The output are portable and efficient HDF5 files, one file per two point
function evaluated, containing the raw Probability Distribution Function (PDF)
of the output values of each function.  The PDFs can then be analyzed or
integrated using several included analysis convenience functions.

\subsection{Time Series Analysis} \label{sec:time_series}

Nearly all of the machinery of \yt{} is designed to operate on a single
simulation dataset, providing spatial analysis while ignoring the time domain.
Time domain analysis is often performed by embedding an analysis script within
a \texttt{for} loop, which requires the user to work out by hand exactly which
datasets from the simulation correspond to the targeted time or redshift
interval.  Matters can be further complicated if a simulation is configured to
output data based on more criteria than simply intervals of constant time.
This is often the case in cosmological simulations where the user will also
specify a list of redshifts at which data should be written.  In the case of
Enzo, redshift and time outputs follow independent naming schemes, making it
difficult to construct a single, time-ordered list of dataset paths.  However,
in much the same way that the parameter file associated with a dataset provides
all of the information necessary to contextualize the contained data, the
parameter file used to initialize a simulation provides everything that is
needed to understand the interrelationship of all the datasets created by that
simulation.

Just as a \yt{} object (called a \texttt{StaticOutput}) is generated for each
distinct simulation output, a \yt{} object called a \texttt{TimeSeries} is
instantiated with the parameter file of the simulation.  Upon initialization,
the time series object will extract from the simulation parameter file all the
information required to know exactly what data has been produced, assuming the
simulation has run to completion.  The time series object contains an ordered
list of datasets corresponding to the time interval to be analyzed.  Much like
data containers such as spheres and regions are created with respect to a
single dataset, they can also be created in relation to the time series object.
These time series data containers can be operated on in the same way as has
already been described, except that in this case any analysis is performed
sequentially on each dataset encompassed by the time series object.  This
enables scripts to address time domain analysis in a much more straightforward
fashion.

\subsection{Synthetic Cosmological Observations}

Conventional techniques for visualizing simulation data capture the state
within the computational domain at one instance in time.  However, true
astronomical observations sample the universe at continually earlier epochs as
they peer further away.  An approximation to this is accomplished by stacking
multiple datasets from different epochs of a single simulation in the line of
sight such that the total comoving radial distance from end to end of the stack
spans the desired redshift interval.  The comoving radial distance ($D_{C}$),
or lines of sight distance, (see \citet{1999astro.ph..5116H} and
\citet{1993ppc..book.....P} for an explanation of this and other cosmological
distance measures) over the redshift range, $z_{1}$ to $z_{2}$, is given by
\begin{equation}
D_{C} = D_{H} \int_{z_{1}}^{z_{2}} \frac{dz}{E(z)},
\end{equation}
where $D_{H} \equiv c/H_{0}$, $c$ is the speed of light, $H_{0}$ is the Hubble
constant, and $E(z)$ is the expansion factor, defined as
\begin{equation}
E(z) \equiv \sqrt{\Omega_{M} (1 + z)^{3} + \Omega_{k} (1 + z)^{2} + 
  \Omega_{\Lambda}}.
\end{equation}
$\Omega_{M}$, $\Omega_{k}$, and $\Omega_{\Lambda}$ are the contributions to the
total energy density of the universe by matter, curvature, and the cosmological
constant.  The user specifies the redshift of the observer and the redshift
interval of the observation.  The time series machinery discussed in
\S\ref{sec:time_series} is used to make the preliminary selection of datasets
appropriate for the requested redshift interval.  The dataset stack is
constructed moving from the upper limit of the redshift interval to the lower
limit.  The dataset at the redshift closest to the upper limit of the requested
interval is the first added to the stack.  Next, the $\delta z$ corresponding
to the length of the simulation box is calculated.  Note that this $\delta z$
is not constant with redshift.  The next dataset added to the stack is chosen
to be the one whose redshift is closest to, but no less than, ($z - \delta z$)
of the last dataset in the stack.  This process continues until the lower limit
of the redshift range is reached.  This method minimizes the number of datasets
required to span a given redshift interval, but the user may specify that a
smaller fraction of the total box length be used to calculate the maximum
$\delta z$ allowable between two datasets in the stack.  

Two different styles of observations can be created from the above
construction: ``light cone'' projections for imaging and ``light rays'' as
synthetic quasar sight lines.  A light cone projection exists as the
combination of projections of each dataset in the stack.  To make a light cone
projection, the user must also specify the angular field of view and resolution
of the image.  As discussed previously, the comoving radial distance determines
the fraction of the box in the line of sight that is used in the projection.
In the plane of the image, the width of the region sample for an image with
field of view, $\theta$, is $[(1 + z)\ D_{A}\ \theta]$, where $D_{A}$ is the
angular size distance, expressed as
\begin{equation}
D_{A} = \frac{1}{1 + z}\left\{
\begin{array}{l l}
D_{H} \frac{1}{\sqrt{\Omega_{k}}} \sinh (\sqrt{\Omega_{k}}
\frac{D_{C}}{D_{H}}) & \quad \mathrm{for}\ \Omega_{k} > 0\\
D_{C}  & \quad \mathrm{for}\ \Omega_{k} = 0\\
D_{H} \frac{1}{\sqrt{|\Omega_{k}|}} \sin (\sqrt{|\Omega_{k}|}
\frac{D_{C}}{D_{H}}) & \quad \mathrm{for}\ \Omega_{k} < 0
\end{array}
\right.
\end{equation}
To minimize the likelihood that the same structures are sampled more than once
along the line of sight, the projection axis and the center of the projected
region are chosen randomly for each dataset in the stack, taking advantage of
the periodicity of the computational domain.  The \yt{} implementation of this
method has been used in \citet{2009ApJ...698.1795H}.  Although this method is
not unique to \yt, e.g. \citet{2007ApJ...671...27H}, certain elements of
\yt{}'s projection machinery (see \S\ref{sec:projections} and
\S\ref{sec:image_creation}) provide great advantages to this implementation.
Though only a fraction of the domain of each dataset in the stack is needed for
projection, the full domain in the lateral directions (but not in the line of
sight) is, in fact, projected.  Since a projection object has been created for
the entire domain, additional light cone projections sampling unique regions of
the domain can be made with no further projection required.  This can greatly
expedite the process of making a large number of light cone projections for
statistical analysis.  \yt{} has the ability to calculate the amount of common
volume sample by two different light cone projections.  A unique solution
generator exists to find a set of random realizations that have a
user-specified maximum common volume.  The unique solution generator will first
attempt to vary only the randomization in the lateral direction, allowing a
single set of projection objects to be used more than once, before attempting a
fully unique randomization.

Light rays rely on the same dataset stack as light cone projections, except the
data object created for each dataset in the stack is an arbitrary-angle ray
instead of a projection.  Just as with light cone projections, each ray segment
has a random orientation for each dataset in the stack.  A ray segment contains
an array of all of the pixels intersected by the ray as well as the path
length, $dl$, of the ray through each pixel.  Therefore, the column density,
$N$, corresponding to a physical density, $\rho$, for an individual length
element of the ray is simply $\rho \times dl$.  Knowing the redshift of each
dataset and the position of a point along the ray, each length element can be
assigned a unique redshift assuming a smooth Hubble flow, allowing for the
creation of synthetic spectra with properly redshifted lines.

\subsection{Level Set Identification}\label{sec:level_sets}

Visual inspection of simulations provides a simple method of identifying
distinct hydrodynamic regions; however, a quantitative approach must be taken
to describe those regions.  Specifically, distinct collapsing regions can be
identified by locating topologically-connected sets of cells.  The nature of
adaptive mesh refinement, wherein a given set of cells may be connected across
grid and refinement boundaries, requires traversing grid and resolution
boundaries.

Unfortunately, while locating connected sets inside a single-resolution grid is
a straightforward but non-trivial problem in recursive programming, extending
this in an efficient way to hierarchical datasets can be problematic.  To that
end, the algorithm implemented in \yt{} checks on a grid-by-grid basis,
utilizing a buffer zone of cells at the grid boundary to join sets that span
grid boundaries.  The algorithm for identifying these sets is a recursive and
iterative process.


\begin{enumerate}
\item Identify grid patches to be considered, such as from a sphere or rectangular
      prism.
\item Give unique identification numbers to all finest-level cells within the
      desired level set $(v_{\mathrm{min}} \leq v \leq v_{\mathrm{max}})$

   \begin{enumerate}
   \item Give unique identification number to all coarse-cells in considered grid
         within desired level set $(v_{\mathrm{min}} \leq v \leq
         v_{\mathrm{max}})$
   \item Obtain buffer zone of one cell-width, including level set IDs
   \item Recursively examine all cells identified as level set members:

      \begin{enumerate}
      \item Update level set ID to be the maximum of 26 neighboring cells
      \item If current level set ID is greater than original level set ID,
            repeat until it is not
      \item Notify all neighboring cells with level set ID less than current
            level set ID to re-examine neighbors and update 
      \end{enumerate}
   \end{enumerate}

\item Construct relationship across grid boundaries.  Any level set
     neighboring the grid boundary is added to a ``join tree,'' composed of
     an old level set ID and a new level set ID.
\item Flatten join tree by ensuring all nodes are unique and do not reference
     any other nodes in the join tree.
\item Coalesce level sets by assigning new level set IDs to those cells that are
     referenced in the join tree.
\item Reorder level set IDs such that the largest level sets have the lowest
      numbers
\item Return extracted level set objects
\end{enumerate}

Once level sets are identified, they are split into individual data containers
(instances of \texttt{ExtractedRegion}) that are returned to the user.  This
presents an integrated interface for generating and analyzing
topologically-connected sets of related cells.  This method was used in
\cite{2009ApJ...691..441S,2009Sci...325..601T} to study fragmentation of
collapsing gas clouds, specifically to examine the gravitational boundedness of
these clouds and the length and density scales at which fragmentation occurs.

To determine whether or not an object is bound, we evaluate the inequality

$$\sum_{i=1}^{N}\frac{m_iv_i^2}{2} <
\sum_{i=1}^{N-1}\sum_{j=i+1}^{N}\frac{Gm_im_j}{r}$$

where $n$ is the number of cells in the identified level set.  The left hand
side of this equation is the total kinetic energy in the object; if desired,
the internal thermal energy ($nkT / (\gamma-1)$) can also be added to this
term.  This code has been written to run either in a hand-coded C module or on
the graphics processor, using NVIDIA's CUDA framework \citep{CUDAGuide2.0} via
the PyCUDA \citep{pycuda_paper} package.  Future versions of the level
set-identification algorithm will implement the method described in
\cite{602127}, which has been designed to be fast and parallelizable.

\section{Future Directions}\label{sec:future_directions}

Development on \yt{} is driven by the pragmatic needs of working astrophysics
researchers.  Several clear goals need to be met in the future, particularly as
the size and scope of simulation data increases.  We also hope to work with
other research groups to add support for the visualization and analysis of
output from other popular astrophysics simulation codes such as ART, Gadget,
Pluto \citep{2007ApJS..170..228M}, and ZEUS-MP \citep{2006ApJS..165..188H}.

The most relevant improvement for very large simulation datasets is to improve
load balancing for parallel operations.  As noted above, for some operations
this can be accomplished by avoiding image-plane decomposition.  Several
efforts are underway to this end. Both the volume rendering and projection
algorithms load balance through decomposition of the image plane, which often
leads to poor work distribution.  These shortcomings are being addressed
algorithmically: adaptive projections will utilize a quad tree, enabling better
load balancing, and volume rendering will utilize a kD-tree approach combined
with intermediate image composition.

However, an underlying problem with the parallelization as it stands is the
global state; each instance of a Python interpreter running \yt{} currently
runs in either ``parallel'' or ``serial'' mode.  Future versions of the \yt{}
parallel analysis interface will allow this to be toggled based on the task
under consideration, as well as more convenience functions for distributing
work tasks between processors--for instance, scatter/gather operations on
halos.

Improvements to the communication mechanisms for parallel analysis in \yt{}
will be necessary as \emph{in situ} analysis grows more pervasive in large
calculations.  \emph{In situ} analysis is challenging as it must necessarily
proceed asynchronously with the simulation; this will require careful data
transport between \yt{} and the simulation code.  Abstracting and isolating the
engine that drives this communication will be necessary to enable \yt{} to be
embedded in simulation codes other than Enzo.

With the widespread deployment of Graphics Processing Units (GPUs) and other
accelerators to supercomputing centers, we will explore using them for fast
numerical computation.  The primary support for GPUs in Python is to enable
dynamic CUDA or OpenCL \citep{CUDAGuide2.0, opencl08} kernel compilation as
well as transparent hosting of arrays in GPU memory \citep[e.g.,
see][]{pycuda_paper}.  These hosted arrays implement many, if not all, NumPy
array operations and could be used as a drop-in replacement in \yt{}.  This
could provide a working mechanism for many numeric calculations to be conducted
in parallel on the GPU.  In order to ensure that offloading computation from
the CPU to the GPU is effective, the entire \yt{} code base will have to be
audited to avoid unnecessary copying of arrays and to ensure that as many array
operations as possible are conducted in-place.  These particular ``hot spots''
provide minimum overhead in standard CPU computing, but could be extremely
detrimental or even cause difficult-to-debug failures in a mixed CPU/GPU
environment.  Furthermore, ensuring that mixed-mode operation is robust and
reliable will be a difficult goal to reach, particularly as \yt{} must support
both CPU and CPU+GPU computation modes.  We are optimistic about exploring
mixed-mode operation in the future, but ensuring its reliability and robustness
will be challenging.  An additional concern is that CUDA is currently a
proprietary standard, and the development of the open standard OpenCL is not as
fast-paced.

At many supercomputing centers, toolkits for constructing graphical user
interfaces are not available or are extremely difficult to build and install.
This greatly reduces the utility of creating a traditional GUI.  To circumvent
this limitation we will be providing a fully-integrated GUI for \yt{} written
in HTML and Javascript and served by a webserver running inside \yt{} itself.
Rather than a large, bulky framework within which operations could be
constructed and executed, this GUI will present a simple interactive
interpreter that can be accessed through a web browser.  This hosted
interpreter would be aware of the hosting web page, and it would dynamically
create user interface widgets as well as enable the inline display of
newly-created images.  As this requires no server-side libraries or widgets,
and as Python itself provides all of the necessary tools on top of which this
type of GUI could be built, we anticipate that this will be far more
maintainable and straightforward than a traditional GUI.  A user will create a
new server on-demand on a supercomputing center login node and connect to it
through an SSH tunnel from a local machine such as a laptop.  Remote analysis
and visualization will be guided and driven through the locally-rendered web
page, with results and images passed back asynchronously and displayed inline
in the same web page.

\section{Conclusions}\label{sec:conclusions}

The \yt{} project is fully free and open source software, with no dependencies
on external code that is not also free and open source software.  The
development process occurs completely in the open at
\url{http://yt.enzotools.org/}, with publicly-accessible source control
systems, bug tracking, mailing lists, and regression tests.  Building a
community of users has been a priority of the \yt{} development team, both to
encourage collaboration and to solicit contributions from new developers; both
the user and developer communities are highly distributed around the world.
\yt{} provides both online and downloadable documentation, covering
introductions to the code, narrative discussion of analysis mechanisms and
modules, and generated documentation covering the data structures and functions
provided by \yt{}'s scripting interface.  A downloadable cookbook provides
scripts for many common end-to-end analysis tasks, all of which provide example
images.

The \yt{} source code comes with a developers' guide, a list of project ideas,
and suggestions for how to get started.  \yt{} is developed using
Mercurial\footnote{\url{http://mercurial.selenic.com/}}, a distributed
version control system that enables local versioned development and encourages
users to make and contribute changes upstream.  A number of additional pieces
of infrastructure assist with community engagement.  \yt{} provides a number of
user-friendly interfaces to assist with debugging, such as a ``pastebin'' that
can be accessed programmatically.  This allows crash reports to be sent
upstream if desired, as well as allowing users to pass around snippets of code
from the command line, mediated by the \yt{} project server.  In the Summer of
2010 a \yt{} users' developers' workshop was held in conjunction with the first
Enzo Users' Workshop at the University of California, San Diego.

In the early days of computational astrophysics, it was common for researchers
to be intimately familiar with the simulation code they used to simulate and
study astrophysical phenomena.  As both computers and simulation codes have
increased in scope and complexity, however, it is now much more common for
groups of researchers to collaborate on the development of a simulation code,
which is then made publicly available and utilized by a much larger community
of less computationally-savvy astrophysicists.  This transition requires the
development of complementary, community-developed analysis and visualization
packages, as well.  We have presented one such analysis package, \yt{},
which is designed to be applicable to multiple simulation codes and to operate
based on physically-relevant quantities.  This abstraction of the underlying
platform enables not only sophisticated examination and visualization of
simulation data, but also cross-code verification and validation of simulation
results.

The creation of a freely available, publicly inspectable platform for
simulation analysis allows the community to disentangle the coding process from
the scientific process.  Simultaneously, by making this platform public,
inspectable and freely available, it can be openly improved and verified.  The
availability and relatively approachable nature of \yt{}, in addition to the
inclusion of many simple analysis tasks, reduces the barrier to entry for young
scientists.  Rather than worrying about the differences between Enzo and FLASH
hierarchy formats, or row versus column ordering, or HDF4 versus HDF5 versus
unformatted fortran data formats, researchers can focus on understanding and
exploring their data.  More generally, however, by orienting the development of
an analysis framework as a community project, the fragmentation of methods and
mechanisms for astrophysical data analysis is greatly inhibited.  Future
generations of simulations and simulation codes will not only benefit from this
collaboration, but they will require it.

\acknowledgments M.J.T.~acknowledges support by NASA ATFP grant NNX08AH26G, NSF
AST-0807312, NSF AST-0807075, as well as the hospitality of the Kavli Institute
for Particle Astrophysics and Cosmology, the Kavli Institute for Theoretical
Physics (during the program ``Star Formation Throughout Cosmic Time'') and the
University of California High-Performance Astro-Computing Center (during the
2010 International Summer School on AstroComputing program).
B.D.S~acknowledges support by NASA grants ATFP NNX09-AD80G and NNZ07-AG77G and
NSF grants AST-0707474 and AST-0908199.  J.S.O.~acknowledges support by NSF
grant AST09-08553.  S.S.~acknowledges support
by NSF grants AST-0708960 and AST-0808184.  S.W.S.~has been supported by NSF
grant AST-0807215 and a DOE Computational Science Graduate Fellowship under
grant number DE-FG02-97ER25308.  We thank Greg Bryan, David Collison,
Ralf Kaehler, Christopher Moody, Brian O'Shea, Joel Primack and John Wise for
their continued support and assistance developing \yt{}.  We would like to
thank Stella Offner, Ji-hoon Kim, John ZuHone, and Oliver Hahn for providing
data, helpful comments, and assistance with Orion, Enzo, FLASH, and RAMSES,
respectively.  We thank the \yt{} community for their support and bug reports
over the last four years.  Additional thanks to Lisandro Dalc\'{\i}n and the
NumPy and Python communities for creating such excellent tools for scientific
analysis.  Portions of \yt{} were developed under the auspices of the National
Nuclear Security Administration of the US Department of Energy at Los Alamos
National Laboratory under Contract No.  DE-AC52-06NA25396, supported by DOE
LDRD grants 20051325PRD4 and 20050031DR.  \yt{} has additionally benefited from
the development supported by the NSF CAREER award AST-0239709 from the National
Science Foundation, The scaling study presented in this work was conducted on
the Triton Resource (\url{http://tritonresource.sdsc.edu/}) at the San Diego
Supercomputer Center.  We also thank the anonymous referee for several helpful
suggestions and comments.


\begin{thebibliography}{52}
\expandafter\ifx\csname natexlab\endcsname\relax\def\natexlab#1{#1}\fi

\bibitem[{{Abel} {et~al.}(2007){Abel}, {Wise}, \&
  {Bryan}}]{2007ApJ...659L..87A}
{Abel}, T., {Wise}, J.~H., \& {Bryan}, G.~L. 2007, \apjl, 659, L87

\bibitem[Agarwal 
\& Feldman(2010)]{2010MNRAS.tmp.1530A} Agarwal, S., \& Feldman, H.~A.\ 2010, \mnras, 1530 

\bibitem[{Ahrens {et~al.}(2005)Ahrens, Geveci, \& Law}]{paraview_paper}
Ahrens, J., Geveci, B., \& Law, C. 2005

\bibitem[{{Almgren} {et~al.}(2010){Almgren}, {Bell}, {Kasen}, {Lijewski},
  {Nonaka}, {Nugent}, {Rendleman}, {Thomas}, \&
  {Zingale}}]{2010arXiv1008.2801A}
{Almgren}, A., {Bell}, J., {Kasen}, D., {Lijewski}, M., {Nonaka}, A., {Nugent},
  P., {Rendleman}, C., {Thomas}, R., \& {Zingale}, M. 2010, ArXiv e-prints
  \url{http://arxiv.org/abs/1008.2801}

\bibitem[{{Bryan} \& {Norman}(1997)}]{bryan97}
{Bryan}, G.~L. \& {Norman}, M.~L. 1997, ArXiv Astrophysics e-prints
\url{http://arxiv.org/abs/astro-ph/9710187}

\bibitem[{{Burns} {et~al.}(2010){Burns}, {Skillman}, \&
  {O'Shea}}]{2010ApJ...721.1105B}
{Burns}, J.~O., {Skillman}, S.~W., \& {O'Shea}, B.~W. 2010, \apj, 721, 1105

\bibitem[Cen \& Ostriker(1992)]{1992ApJ...399L.113C} Cen, R., \& Ostriker,
J.~P.\ 1992, \apjl, 399, L113 

\bibitem[{{Collins} {et~al.}(2010){Collins}, {Padoan}, {Norman}, \&
  {Xu}}]{2010arXiv1008.2402C}
{Collins}, D.~C., {Padoan}, P., {Norman}, M.~L., \& {Xu}, H. 2010, ArXiv
  e-prints \url{http://arxiv.org/abs/1008.2402}

\bibitem[{Dalc\'{\i}n {et~al.}(2008)Dalc\'{\i}n, Paz, \&
  D'Elia}]{MPI4PY:PAPER2}
Dalc\'{\i}n, L., Paz, R., \& D'Elia, M. S.~J. 2008, Journal of Parallel and
  Distributed Computing, 68, 655

\bibitem[{Dalc\'{\i}n {et~al.}(2005)Dalc\'{\i}n, Paz, \&
  Storti}]{MPI4PY:PAPER1}
Dalc\'{\i}n, L., Paz, R., \& Storti, M. 2005, Journal of Parallel and
  Distributed Computing, 65, 1108

\bibitem[{{Eisenstein} \& {Hut}(1998)}]{eishut98}
{Eisenstein}, D.~J. \& {Hut}, P. 1998, \apj, 498, 137

\bibitem[{Forum(1994)}]{MPIStandard}
Forum, M.~P. 1994, MPI: A Message-Passing Interface Standard, Tech. rep., MPI
  Forum, Knoxville, TN, USA

\bibitem[{{Fryxell} {et~al.}(2000){Fryxell}, {Olson}, {Ricker}, {Timmes},
  {Zingale}, {Lamb}, {MacNeice}, {Rosner}, {Truran}, \&
  {Tufo}}]{2000ApJS..131..273F}
{Fryxell}, B., {Olson}, K., {Ricker}, P., {Timmes}, F.~X., {Zingale}, M.,
  {Lamb}, D.~Q., {MacNeice}, P., {Rosner}, R., {Truran}, J.~W., \& {Tufo}, H.
  2000, \apjs, 131, 273

\bibitem[{{Hallman} {et~al.}(2007){Hallman}, {O'Shea}, {Burns}, {Norman},
  {Harkness}, \& {Wagner}}]{2007ApJ...671...27H}
{Hallman}, E.~J., {O'Shea}, B.~W., {Burns}, J.~O., {Norman}, M.~L., {Harkness},
  R., \& {Wagner}, R. 2007, \apj, 671, 27

\bibitem[{{Hallman} {et~al.}(2009){Hallman}, {O'Shea}, {Smith}, {Burns}, \&
  {Norman}}]{2009ApJ...698.1795H}
{Hallman}, E.~J., {O'Shea}, B.~W., {Smith}, B.~D., {Burns}, J.~O., \& {Norman},
  M.~L. 2009, \apj, 698, 1795

\bibitem[{{Hayes} {et~al.}(2006){Hayes}, {Norman}, {Fiedler}, {Bordner}, {Li},
  {Clark}, {ud-Doula}, \& {Mac Low}}]{2006ApJS..165..188H}
{Hayes}, J.~C., {Norman}, M.~L., {Fiedler}, R.~A., {Bordner}, J.~O., {Li},
  P.~S., {Clark}, S.~E., {ud-Doula}, A., \& {Mac Low}, M. 2006, \apjs, 165, 188

\bibitem[{{Hogg}(1999)}]{1999astro.ph..5116H}
{Hogg}, D.~W. 1999, ArXiv Astrophysics e-prints
\url{http://arxiv.org/abs/astro-ph/9905116}

\bibitem[{Hunter(2007)}]{matplotlib_paper}
Hunter, J.~D. 2007, Computing in Science \& Engineering, 9, 90

\bibitem[{{Joye} \& {Mandel}(2003)}]{2003ASPC..295..489J}
{Joye}, W.~A. \& {Mandel}, E. 2003, in Astronomical Society of the Pacific
  Conference Series, Vol. 295, Astronomical Data Analysis Software and Systems
  XII, ed. {H.~E.~Payne, R.~I.~Jedrzejewski, \& R.~N.~Hook}, 489--+

\bibitem[{Kaehler {et~al.}(2005)Kaehler, Prohaska, Hutanu, \&
  Hege}]{10.1109/VIS.2005.122}
Kaehler, R., Prohaska, S., Hutanu, A., \& Hege, H.-C. 2005, Visualization
  Conference, IEEE, 0, 23

\bibitem[{Kaehler {et~al.}(2006)Kaehler, Wise, Abel, \& Hege}]{vg06-kaehler}
Kaehler, R., Wise, J., Abel, T., \& Hege, H.-C. 2006, in {Proceedings of the
  International Workshop on Volume Graphics 2006} (Boston: Eurographics / IEEE
  VGTC 2006), 103--110

\bibitem[{{Khronos OpenCL Working Group}(2008)}]{opencl08}
{Khronos OpenCL Working Group}. 2008, The OpenCL Specification, version 1.0.29

\bibitem[{{Kim} {et~al.}(2009){Kim}, {Wise}, \& {Abel}}]{2009ApJ...694L.123K}
{Kim}, J., {Wise}, J.~H., \& {Abel}, T. 2009, \apjl, 694, L123

\bibitem[{{Klein} {et~al.}(1999){Klein}, {Fisher}, {McKee}, \&
  {Truelove}}]{1999ASSL..240..131K}
{Klein}, R.~I., {Fisher}, R.~T., {McKee}, C.~F., \& {Truelove}, J.~K. 1999, in
  Astrophysics and Space Science Library, Vol. 240, Numerical Astrophysics, ed.
  {S.~M.~Miyama, K.~Tomisaka, \& T.~Hanawa}, 131--+

\bibitem[{Kl\"{o}ckner {et~al.}(2009)Kl\"{o}ckner, Pinto, Lee, Catanzaro,
  Ivanov, \& Fasih}]{pycuda_paper}
Kl\"{o}ckner, A., Pinto, N., Lee, Y., Catanzaro, B., Ivanov, P., \& Fasih, A.
  2009

\bibitem[{{Klypin} {et~al.}(2010){Klypin}, {Trujillo-Gomez}, \&
  {Primack}}]{2010arXiv1002.3660K}
{Klypin}, A., {Trujillo-Gomez}, S., \& {Primack}, J. 2010, ArXiv e-prints
\url{http://arxiv.org/abs/1002.3660}

\bibitem[{{Kravtsov} {et~al.}(1997){Kravtsov}, {Klypin}, \&
  {Khokhlov}}]{1997ApJS..111...73K}
{Kravtsov}, A.~V., {Klypin}, A.~A., \& {Khokhlov}, A.~M. 1997, \apjs, 111, 73

\bibitem[{{Kritsuk} {et~al.}(2007){Kritsuk}, {Norman}, {Padoan}, \&
  {Wagner}}]{2007ApJ...665..416K}
{Kritsuk}, A.~G., {Norman}, M.~L., {Padoan}, P., \& {Wagner}, R. 2007, \apj,
  665, 416

\bibitem[{{Krumholz}(2010)}]{2010arXiv1008.4368K}
{Krumholz}, M.~R. 2010, ArXiv e-prints \url{http://arxiv.org/abs/1008.4368}

\bibitem[{{Krumholz} {et~al.}(2007){Krumholz}, {Klein}, {McKee}, \&
  {Bolstad}}]{2007ApJ...667..626K}
{Krumholz}, M.~R., {Klein}, R.~I., {McKee}, C.~F., \& {Bolstad}, J. 2007, \apj,
  667, 626

\bibitem[{{Krumholz} {et~al.}(2004){Krumholz}, {McKee}, \&
  {Klein}}]{2004ApJ...611..399K}
{Krumholz}, M.~R., {McKee}, C.~F., \& {Klein}, R.~I. 2004, \apj, 611, 399

\bibitem[{{Kuhlen} {et~al.}(2008){Kuhlen}, {Diemand}, {Madau}, \&
  {Zemp}}]{2008JPhCS.125a2008K}
{Kuhlen}, M., {Diemand}, J., {Madau}, P., \& {Zemp}, M. 2008, Journal of
  Physics Conference Series, 125, 012008

\bibitem[{LLNL(2010)}]{chombo_website}
LLNL, A. N. A.~G. 2010, Chombo---Infrastructure for Adaptive Mesh Refinement,
  \texttt{https://seesar.lbl.gov/anag/chombo/index.html}

\bibitem[{{Mignone} {et~al.}(2007){Mignone}, {Bodo}, {Massaglia}, {Matsakos},
  {Tesileanu}, {Zanni}, \& {Ferrari}}]{2007ApJS..170..228M}
{Mignone}, A., {Bodo}, G., {Massaglia}, S., {Matsakos}, T., {Tesileanu}, O.,
  {Zanni}, C., \& {Ferrari}, A. 2007, \apjs, 170, 228

\bibitem[{{Norman} {et~al.}(2007){Norman}, {Bryan}, {Harkness}, {Bordner},
  {Reynolds}, {O'Shea}, \& {Wagner}}]{2007arXiv0705.1556N}
{Norman}, M.~L., {Bryan}, G.~L., {Harkness}, R., {Bordner}, J., {Reynolds}, D.,
  {O'Shea}, B., \& {Wagner}, R. 2007, ArXiv e-prints
  \url{http://arxiv.org/abs/0705.1556}

\bibitem[{{NVIDIA}(2008)}]{CUDAGuide2.0}
{NVIDIA}. 2008, NVIDIA CUDA Programming Guide 2.0

\bibitem[{{Ocvirk} {et~al.}(2008){Ocvirk}, {Pichon}, \&
  {Teyssier}}]{2008MNRAS.390.1326O}
{Ocvirk}, P., {Pichon}, C., \& {Teyssier}, R. 2008, \mnras, 390, 1326

\bibitem[Offner et al.(2008)]{2008Offnerposter} Offner, S., Klein, R. I.,
McKee, C.~F., \& Chakrabarti, S.\ 2008, "Radiative Transfer Simulations:
Low-Mass Cores, Disks, and Protostars", New Light on Young Stars: Spitzer's
View of Circumstellar Disks, Pasadena, CA.

\bibitem[{Oliphant(2007)}]{numpy_paper}
Oliphant, T.~E. 2007, Computing in Science and Engineering, 9, 10

\bibitem[{{O'Shea} {et~al.}(2004){O'Shea}, {Bryan}, {Bordner}, {Norman},
  {Abel}, {Harkness}, \& {Kritsuk}}]{oshea04}
{O'Shea}, B.~W., {Bryan}, G., {Bordner}, J., {Norman}, M.~L., {Abel}, T.,
  {Harkness}, R., \& {Kritsuk}, A. 2004, ArXiv Astrophysics e-prints
  \url{http://arxiv.org/abs/astro-ph/0403044}

\bibitem[{Pascucci \& Cole-McLaughlin(2002)}]{602127}
Pascucci, V. \& Cole-McLaughlin, K. 2002, in VIS '02: Proceedings of the
  conference on Visualization '02 (Washington, DC, USA: IEEE Computer Society),
  187--194

\bibitem[{{Peebles}(1993)}]{1993ppc..book.....P}
{Peebles}, P.~J.~E. 1993, {Principles of physical cosmology}, ed. {Peebles,
  P.~J.~E.}

\bibitem[{{Reynolds} {et~al.}(2009){Reynolds}, {Hayes}, {Paschos}, \&
  {Norman}}]{2009JCoPh.228.6833R}
{Reynolds}, D.~R., {Hayes}, J.~C., {Paschos}, P., \& {Norman}, M.~L. 2009,
  Journal of Computational Physics, 228, 6833

\bibitem[{{Silvia} {et~al.}(2010){Silvia}, {Smith}, \&
  {Shull}}]{2010ApJ...715.1575S}
{Silvia}, D.~W., {Smith}, B.~D., \& {Shull}, J.~M. 2010, \apj, 715, 1575

\bibitem[{{Skillman} {et~al.}(2010){Skillman}, {Hallman}, {O'Shea}, {Burns},
  {Smith}, \& {Turk}}]{2010arXiv1006.3559S}
{Skillman}, S.~W., {Hallman}, E.~J., {O'Shea}, B.~W., {Burns}, J.~O., {Smith},
  B.~D., \& {Turk}, M.~J. 2010, ArXiv e-prints
  \url{http://arxiv.org/abs/1006.3559}

\bibitem[{{Skory} {et~al.}(2010){Skory}, {Turk}, {Norman}, \&
  {Coil}}]{2010arXiv1001.3411S}
{Skory}, S., {Turk}, M.~J., {Norman}, M.~L., \& {Coil}, A.~L. 2010, ArXiv
  e-prints \url{http://arxiv.org/abs/1001.3411}

\bibitem[{{Smith} {et~al.}(2008){Smith}, {Sigurdsson}, \&
  {Abel}}]{2008MNRAS.385.1443S}
{Smith}, B., {Sigurdsson}, S., \& {Abel}, T. 2008, \mnras, 385, 1443

\bibitem[{{Smith} {et~al.}(2009){Smith}, {Turk}, {Sigurdsson}, {O'Shea}, \&
  {Norman}}]{2009ApJ...691..441S}
{Smith}, B.~D., {Turk}, M.~J., {Sigurdsson}, S., {O'Shea}, B.~W., \& {Norman},
  M.~L. 2009, \apj, 691, 441

\bibitem[{{Springel}(2005)}]{2005MNRAS.364.1105S}
{Springel}, V. 2005, \mnras, 364, 1105

\bibitem[{{Springel} {et~al.}(2005){Springel}, {White}, {Jenkins}, {Frenk},
  {Yoshida}, {Gao}, {Navarro}, {Thacker}, {Croton}, {Helly}, {Peacock}, {Cole},
  {Thomas}, {Couchman}, {Evrard}, {Colberg}, \& {Pearce}}]{2005Natur.435..629S}
{Springel}, V., {White}, S.~D.~M., {Jenkins}, A., {Frenk}, C.~S., {Yoshida},
  N., {Gao}, L., {Navarro}, J., {Thacker}, R., {Croton}, D., {Helly}, J.,
  {Peacock}, J.~A., {Cole}, S., {Thomas}, P., {Couchman}, H., {Evrard}, A.,
  {Colberg}, J., \& {Pearce}, F. 2005, \nat, 435, 629

\bibitem[{{Teyssier}(2002)}]{2002A&A...385..337T}
{Teyssier}, R. 2002, \aap, 385, 337

\bibitem[{{Truelove} {et~al.}(1998){Truelove}, {Klein}, {McKee}, {Holliman},
  {Howell}, {Greenough}, \& {Woods}}]{1998ApJ...495..821T}
{Truelove}, J.~K., {Klein}, R.~I., {McKee}, C.~F., {Holliman}, II, J.~H.,
  {Howell}, L.~H., {Greenough}, J.~A., \& {Woods}, D.~T. 1998, \apj, 495, 821

\bibitem[{{Turk} {et~al.}(2009){Turk}, {Abel}, \&
  {O'Shea}}]{2009Sci...325..601T}
{Turk}, M.~J., {Abel}, T., \& {O'Shea}, B. 2009, Science, 325, 601

\bibitem[{Weber {et~al.}(2010)Weber, Ahern, Bethel, Borovikov, Childs, Deines,
  Garth, Hagen, Hamann, Joy, Martin, Meredith, Prabhat, Pugmire, R{\"u}bel,
  Van~Straalen, \& Wu}]{visit_paper}
Weber, G.~H., Ahern, S., Bethel, E.~W., Borovikov, S., Childs, H.~R., Deines,
  E., Garth, C., Hagen, H., Hamann, B., Joy, K.~I., Martin, D., Meredith, J.,
  Prabhat, Pugmire, D., R{\"u}bel, O., Van~Straalen, B., \& Wu, K. 2010, in
  Numerical Modeling of Space Plasma Flows: Astronum-2009 (Astronomical Society
  of the Pacific Conference Series), lBNL-3185E. To appear

\end{thebibliography}
\end{document}